\documentclass[twocolumn,preprintnumbers,amsmath,amssymb,showpacs]{revtex4}
\usepackage{epsfig}
\begin{document}
\setlength{\voffset}{1.0cm}
\title{Phase structure of the massive chiral Gross-Neveu model from Hartree-Fock}
\preprint{FAU-TP3-08/05}
\author{Christian Boehmer\footnote{christian.boehmer@theorie3.physik.uni-erlangen.de}}
\author{Ulf Fritsch\footnote{ulf.fritsch@physik.uni-erlangen.de}}
\author{Sebastian Kraus}
\author{Michael Thies\footnote{thies@theorie3.physik.uni-erlangen.de}}
\affiliation{Institut f\"ur Theoretische Physik III,
Universit\"at Erlangen-N\"urnberg, D-91058 Erlangen, Germany}
\date{\today}
\begin{abstract}
The phase diagram of the massive chiral Gross-Neveu model (the massive Nambu--Jona-Lasinio model in 1+1 dimensions)
is constructed. In the large $N$ limit, the Hartree-Fock approach can be used. We find numerically a chiral crystal phase
separated from a massive Fermi gas phase by a 1st order transition. Using perturbation theory, we also construct the critical
sheet where the homogeneous phase becomes unstable in a 2nd order transition. A tricritical curve is located. The phase
diagram is mapped out as a function of fermion mass, chemical potential and temperature and
compared with the one of the discrete chiral Gross-Neveu model. As a by-product, we illustrate the crystal structure of
matter at zero temperature for various densities and fermion masses. 
\end{abstract}
\maketitle

\section{Introduction}\label{sect1}

To map out the phase diagram of hot and dense matter has been a major goal of strong interaction physics during
the last decades, both experimentally and theoretically. As is often the case, these efforts have been accompanied by 
studies of drastically simplified, solvable model problems to sharpen the theoretical tools and get guidance for 
more realistic cases. Among the few known field theories which are both solvable and possess a non-trivial
phase structure, fermionic large $N$ models in 1+1 dimensions like the 't Hooft model \cite{R1} or Gross-Neveu models \cite{R2} 
are perhaps most instructive, as they share a number of properties with quantum chromodynamics (for a pedagogical 
review, see Ref.~\cite{R3}). Given that these models have been formulated back in 1974 already,
it is surprising that their phase diagrams as a function of temperature, chemical potential
and fermion mass have not yet been fully established. As far as we can tell, the reason is not that these phase 
diagrams were considered to be uninteresting. Rather, this situation reflects a shortcoming of the first round of theoretical
investigations during the 80's and 90's with methods too crude to expose the full, rich phase structure. As a consequence, there
has been renewed interest recently in this topic with results which also have some bearing on low dimensional condensed
matter systems, and the original phase diagrams are still in the process of revision right now. For an update on the current state of the art, 
see Refs. \cite{R4,R5} and references therein.

In the present paper, we focus on the phase structure of the massive, chiral Gross-Neveu (GN) model at finite temperature and chemical 
potential. This model is nothing but the 1+1 dimensional Nambu--Jona-Lasinio model \cite{R6} with $N$ fermion flavors and a 
bare mass term explicitly breaking the U(1)$\otimes$U(1) chiral symmetry. Its Lagrangian reads 
\begin{equation}
{\cal L}= \bar{\psi} ({\rm i}\partial \!\!\!/-m_0)\psi + \frac{g^2}{2}\left[ (\bar{\psi}\psi)^2+(\bar{\psi}{\rm i}\gamma_5\psi)^2\right]
\label{A1}
\end{equation}
where flavor indices are suppressed as usual, i.e., $\bar{\psi}\psi=\sum_{k=1}^N \bar{\psi}_k \psi_k$ etc. 
We are only interested in the 't~Hooft limit ($N\to \infty, Ng^2={\rm const.}$) in which classic no-go theorems can be
bypassed and breakdown of continuous symmetries becomes possible in 1+1 dimensions.
In spite of the fact that semiclassical methods for solving such models have been developed in the 70's already \cite{R2,R7},  
the full phase diagram of the simple field theory with Lagrangian (\ref{A1}) is still largely unknown.
Consider first the chiral limit ($m_0=0$) of the model. If one constrains the condensates $\langle \bar{\psi}\psi \rangle,
\langle \bar{\psi}{\rm i}\gamma_5 \psi \rangle$ to be spatially constant, the resulting phase diagram is identical to the one
from the simpler GN model variant with discrete chiral symmetry and scalar-scalar coupling $(\bar{\psi}\psi)^2$ only.
One finds two phases in the ($\mu,T$) plane, a massless and a massive Fermi gas, separated by 
1st and 2nd order transitions \cite{R8}. As soon as one allows for spatially  inhomogeneous condensates,
the system takes advantage of the Peierls effect \cite{R9} and opens a gap at the Fermi surface. 
This results in a solitonic chiral crystal phase. The first crystalline solution of the Hartree-Fock (HF) problem which was 
found is the ``chiral spiral" with helical order parameter and a strikingly different phase diagram \cite{R10,R3}.
As pointed out in Ref.~\cite{R10a}, these results can also be understood readily in terms of bosonization.
Very recently however, they have been challenged by a more sophisticated candidate for the complex order parameter 
in the form of a chirally twisted crystal, using powerful resolvent methods to 
generate self-consistent solutions in closed analytical form \cite{R11,R5}. The implications for the phase diagram have not yet
been fully worked out but promise an even richer structure of the solitonic crystal phase than previously thought.   

Turning to the massive chiral GN model ($m_0>0$), we first should like to remind the reader that the bare parameters $g^2,m_0$
in Eq.~(\ref{A1}) together with the UV cutoff $\Lambda/2$ get replaced by two physical, renormalization group invariant 
parameters $m$ and $\gamma$ in the process of regularization and renormalization \cite{R4,R10b}. Here, $m$ is the physical fermion mass
in the vacuum (set equal to 1 without loss of generality throughout this paper) and $\gamma$ the ``confinement parameter" measuring the
explicit violation of chiral symmetry, 
\begin{equation}
\frac{\pi}{Ng^2}= \gamma + \ln \frac{\Lambda}{m}, \qquad \gamma:= \frac{\pi}{Ng^2}\frac{m_0}{m}.
\label{A9}
\end{equation}
The following bits and pieces are known about the phase structure of the massive model.
The phase diagram assuming $x$-independent condensates only \cite{R12} is once again indistinguishable from that of the massive discrete
chiral GN model, but this is an artefact of the assumption of homogeneity \cite{R4}. As far as inhomogeneous condensates are concerned, 
it is useful to start from the low density, low temperature limit governed by the isolated baryons of the model.
Baryons of the massive chiral GN model were first studied near the chiral limit by means of variational techniques \cite{R13} and subsequently
via the 
derivative expansion \cite{R14}. They turn out to be closely related to the sine-Gordon kink. A recent numerical HF calculation,
supplemented by 
analytical asymptotic expansions, has been able to follow the baryon mass and structure to arbitrary $\gamma$ \cite{R15}.
This was actually done in preparation of the present study. Unlike in the discrete chiral GN model, the self-consistent baryon potentials found
were not reflectionless, 
a serious obstacle for a full analytical solution. Aside from individual baryons relevant to the base line at $T=0$ 
of the ($\gamma,\mu,T$) phase diagram, the vicinity of the tricritical point ($\gamma=0,\mu=0,T={\rm e}^{\rm C}/\pi$) has also been explored
in some detail \cite{R16}.
The phase structure was deduced from a microscopic Ginzburg-Landau (GL) approach, based once again on the derivative expansion.
In this work, both first and second order critical lines between homogeneous and inhomogeneous phases were identified.
As a result, one already starts to see that the GN models with (broken) discrete and continuous chiral symmetry have totally
different phase diagrams, as is indeed expected on the basis of universality arguments. 

In the present paper, we report on a solution of the HF problem at finite $T,\mu$ for a whole range of $\gamma$ values and 
construct a first candidate for the full phase diagram of the massive chiral GN model. It is not yet known what impact the 
more general, chirally twisted soliton crystals of the massless model discovered in Ref.~\cite{R11} would have on the massive model, and 
we cannot contribute anything to this question. 
Our aim here is to extend the calculations of Ref.~\cite{R16} near the tricritical point to a significant portion of ($\mu,T,\gamma$)
space, so that a 3d plot of the phase diagram can be drawn and compared with the one from the discrete chiral GN model. We think that 
such an undertaking  is worthwhile in the present situation, but should be followed up by efforts to identify alternative chiral crystal
structures which might be thermodynamically more stable \cite{R11}, or by further attempts to arrive at a full analytical solution as in the
case of the discrete chiral GN model \cite{R4}.

The remaining paper is organized as follows. Sec.~\ref{sect2} is devoted to the HF calculation at zero temperature. We explain the
general numerical 
procedure (\ref{sect2a}), discuss analytically the low and high density asymptotics (\ref{sect2b}) and present selected numerical results
(\ref{sect2c}).
Sec.~\ref{sect3} contains all the material about finite temperature and the phase diagram. We briefly outline the thermal HF approach
to the grand canonical potential (\ref{sect3a}) and recall previous results from GL theory (\ref{sect3b}). In Sec.~\ref{sect3c},
we describe in detail how we have obtained the perturbative, 2nd order critical sheet. The non-perturbative 1st order sheet represents the 
most difficult part of our analysis, since we can only determine it numerically at present. This is presented in Sec.~\ref{sect3d} along with the  
final results. In the concluding Sec.~\ref{sect4}, we summarize our findings, compare the phase diagram with other related phase
diagrams and identify areas where more work is needed. 

\section{Hartree-Fock calculation of dense matter at $T=0$}\label{sect2}

\subsection{Setup of the numerical calculations}\label{sect2a}

The HF calculation in the chiral GN model starts from the Dirac Hamiltonian
\begin{equation}
H= \gamma_5 \frac{1}{\rm i} \frac{\partial}{\partial x}+ \gamma^0 S(x) +{\rm i}\gamma^1 P(x)
\label{A2}
\end{equation}
with scalar and pseudoscalar potentials $S,P$ to be determined self-consistently. 
In Ref.~\cite{R15}, a numerical HF study including the Dirac sea has been used to construct the baryons
of this model. For technical reasons, the calculation was done in a finite interval of length $L$ with
antiperiodic boundary conditions for the fermions, using a basis of free, massive spinors in discretized momentum space. 
Now assume that $S,P$ are periodic with spatial period $a$. We can actually reduce the HF calculation for 
such a crystal to the one for a single baryon performed in \cite{R15}. We enclose the crystal in a box of length $L=N a$ 
containing $N$ periods and impose again antiperiodic boundary conditions on the fermion single particle
wave functions in this large interval,
\begin{equation}
\psi(L)=-\psi(0).
\label{A3}
\end{equation}
According to the Bloch theorem, the eigenspinors of $H$ are of the form
\begin{equation} 
\psi(x)=\phi(x){\rm e}^{{\rm i}px}, \qquad \phi(x+a)=\phi(x).
\label{A4}
\end{equation}
The boundary condition (\ref{A3}) discretizes the Bloch momenta,
\begin{equation}
p_n=\frac{2\pi}{L}\left( n+ \frac{1}{2} \right), \qquad (n\in \mathbb{Z}).
\label{A5}
\end{equation}
For a single period, e.g., the interval $[0,a]$, this implies quasi-periodic boundary conditions,
\begin{equation}
\psi(a) = {\rm e}^{{\rm i} \beta_{\nu}}\psi(0), 
\label{A6}
\end{equation}
where the $N$ discrete values of $\beta_{\nu}$ parametrize the $N$-th roots of $(-1)$,
\begin{equation}
\beta_{\nu} = \frac{2\pi}{N}\left(\nu + \frac{1}{2}\right), \qquad \nu=0,1,...,N-1.
\label{A7}
\end{equation}
Hence, to get the spectrum of $H$ with a periodic potential, all we have to do is compute the spectrum for a single ``baryon" 
in an interval of length $a$ with quasi-periodic boundary conditions along the lines of Ref.~\cite{R15}, repeat the calculation
$N$ times (for all possible values of the phase $\beta_{\nu}$) and collect the spectra. This enables us to take over the
calculational method literally from Ref.~\cite{R15}. We also stick to the conditions
\begin{equation}
S(x)=S(-x), \qquad P(x)=-P(-x),
\label{A8}
\end{equation}
reflecting the difference between scalar and pseudoscalar potentials if parity is unbroken.
To evaluate the energy density of the crystal at $T=0$, we once again combine a numerical diagonalization with
perturbation theory for states deep down in the Dirac sea. The technical details like vacuum subtraction, double counting correction
and renormalization are identical to those given in Ref.~\cite{R15} and need not be repeated here.

A key element of the HF approach is self-consistency of the potentials $S,P$. As explained in Ref.~\cite{R15} this can be achieved by
minimizing the HF energy at fixed fermion number with respect to the potentials, provided one varies the potentials without any bias.
In the present work, we assume periodicity, expand $S$ and $P$ into Fourier series,
\begin{equation}
S(x)=\sum_{\ell}S_{\ell}{\rm e}^{{\rm i}2\pi\ell x/a}, \quad P(x)={\rm i}\sum_{\ell}P_{\ell}{\rm e}^{{\rm i}2\pi\ell x/a},
\label{A10}
\end{equation}
and minimize the HF energy with respect to the Fourier coefficients $S_{\ell},P_{\ell}$ and the spatial period $a$, using a standard
conjugate gradient algorithm. The only other bias put in aside from periodicity are the symmetry relations (\ref{A8}).
If the true self-consistent potential $\Delta=S-{\rm i}P$ would not be strictly periodic but carry a chiral twist,
\begin{equation}
\Delta(x+a)={\rm e}^{2{\rm i}\varphi} \Delta(x),
\label{A10a}
\end{equation}
as proposed in a recent study of the massless chiral GN model \cite{R11,R5}, our calculation might still be useful as a variational
calculation, but we could miss the true self-consistent potential. Note however that there is so far no claim 
of non-periodic potentials in the massive model considered in the present work. 

\subsection{Low and high density limits}\label{sect2b}

In the limits of low and high fermion density, the ground state energy can be calculated analytically. 
If the valence band is completely filled (as is indeed found in the full HF calculation), the  
spatially averaged baryon density per flavor is related to the period $a$ via
\begin{equation}
\rho= \frac{1}{a}= \frac{p_f}{\pi}.
\label{A11}
\end{equation}
The last equation defines the Fermi momentum $p_f$.
At very low density, we expect the energy density to be determined by the baryon mass,
\begin{equation}
{\cal E}_{\rm HF}-{\cal E}_{\rm vac} \approx M_{\rm B}\rho 
\label{A12}
\end{equation}
The baryon mass is known already from Ref.~\cite{R15}. 
At high density on the other hand, we can use perturbation theory to predict the asymptotic behaviour of the energy density.
This is a simple generalization of a similar calculation done in Ref.~\cite{R17} for the massive GN model with broken 
discrete chiral symmetry, cf. Eqs.~(67)--(74)  of that paper. It is sufficient to keep the Fourier amplitudes $S_0,S_1$ and $P_1$
for this purpose.
Standard 2nd order perturbation theory then yields the single particle energies
\begin{equation}
E_{\eta,p} = \eta \, {\rm sgn}(p) \left( p + \frac{S_0^2}{2p} + \frac{(S_1+ f P_1)^2}{2(p+p_f)} +\frac{(S_1- f P_1)^2}{2(p-p_f)}
\right)
\label{A13}
\end{equation}
where $\eta=\pm 1$ and
\begin{equation}
f=1-2 \delta_{\eta,{\rm sgn}(p)}.
\label{A14}
\end{equation}
Along the lines of Ref.~\cite{R17}, we find for the perturbative ground state energy
\begin{eqnarray}
{\cal E}_{\rm HF} &=& - \frac{\Lambda^2}{8\pi} + \frac{p_f^2}{2\pi} + \frac{S_0^2}{2\pi}[\gamma + \ln (2p_f)]-\frac{\gamma S_0}{\pi}
\label{A15} \\
&+ &  \frac{y^2}{4\pi}[2 \gamma -1 + \ln(y^2)] +\frac{2X^2}{\pi}[\gamma+ \ln (4p_f)]
\nonumber
\end{eqnarray}
where we have set 
\begin{equation}
S_1=X+y/2, \qquad P_1=X-y/2.
\label{A16}
\end{equation}
Minimizing ${\cal E}_{\rm HF}$ with respect to $S_0, X$ and $y$ yields  
\begin{equation}
X=0, \qquad S_0= \frac{\gamma}{\gamma+ \ln(2p_f)}
\label{A17}
\end{equation}
and the equation
\begin{equation}
y[2\gamma + \ln(y^2)]=0
\label{A18}
\end{equation}
with the solutions $y=0$ (homogeneous condensate) and 
\begin{equation}
y=\pm  {\rm e}^{-\gamma}.
\label{A19}
\end{equation}
The self-consistent potential $\Delta=S-{\rm i}P$ for the non-trivial solution $y={\rm e}^{-\gamma}$ is inhomogeneous,
\begin{equation}
\Delta(x)  =  \frac{\gamma}{\gamma + \ln (2p_f)} + {\rm exp}\left\{ -2{\rm i}p_f x -\gamma\right\}.
\label{A20}
\end{equation}
(The other sign of $y$ merely corresponds to a translation of the crystal by $a/2$.)
The ground state energy (\ref{A15}) at the minimum is indeed lower than the one of the homogeneous solution,   
\begin{equation}
{\cal E}_{\rm HF}(y\neq 0)-{\cal E}_{\rm HF}(y=0)=-\frac{1}{4\pi} {\rm e}^{-2\gamma}.
\label{A21}
\end{equation}
At $\gamma=0$ this agrees with the result for the chiral spiral \cite{R10}.
Finally we write down the ground state energy for large $p_f$, relative to the vacuum. It has the asymptotic behavior
\begin{eqnarray}
{\cal E}_{\rm HF}-{\cal E}_{\rm vac} & \approx &  \frac{p_f^2}{2\pi} - \frac{\gamma^2}{2\pi(\gamma+ \ln 2p_f)}
\nonumber \\ &  & + \frac{1}{4\pi}\left(1+2 \gamma - {\rm e}^{-2\gamma}\right).
\label{A22}
\end{eqnarray}
Eqs.~(\ref{A12}) for $p_f\to 0$ and (\ref{A22}) for $p_f \to \infty$ are the main results of this section, ready to be compared to
full numerical results below. 
\begin{figure}
\begin{center}
\epsfig{file=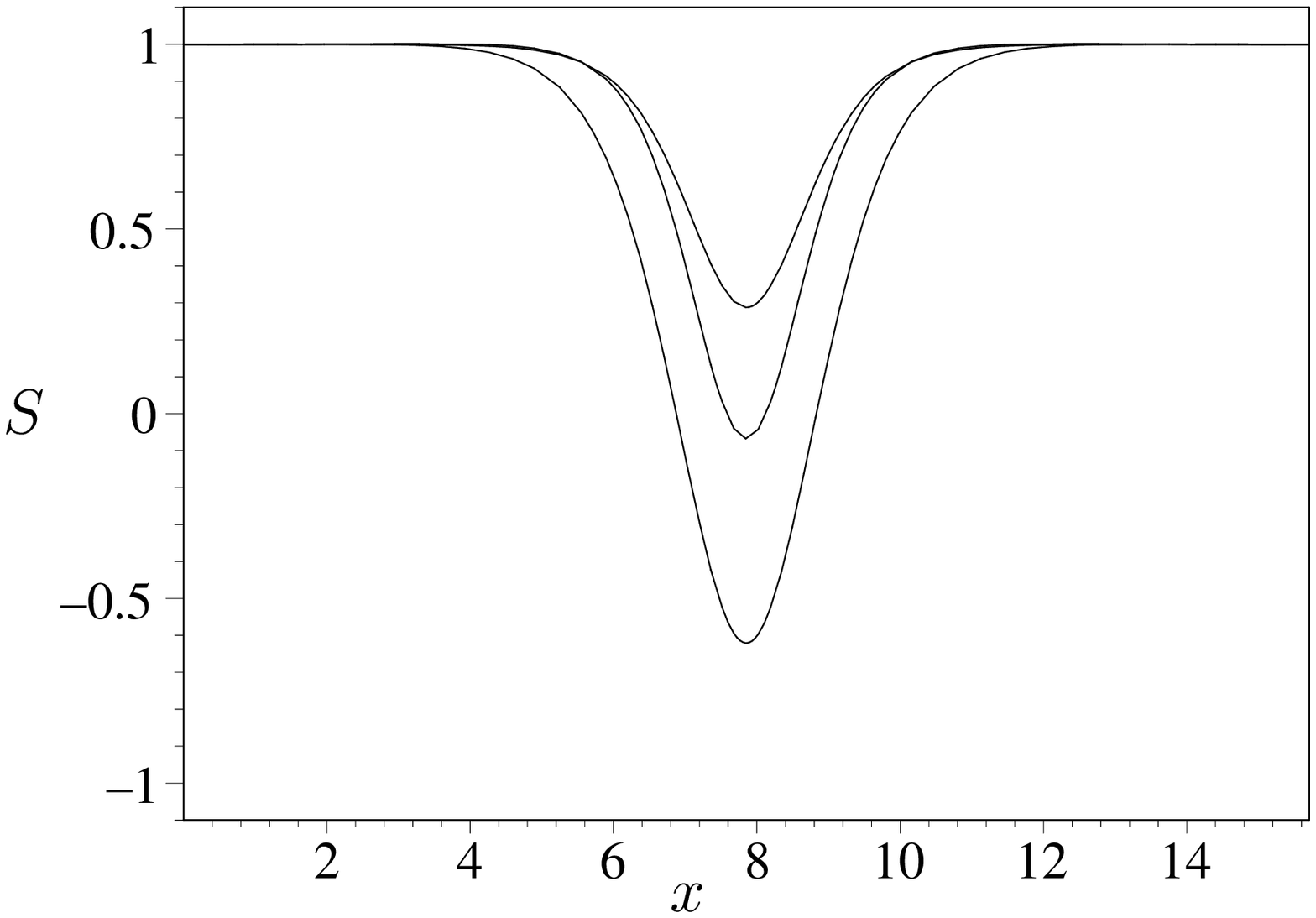,width=6.0cm,angle=270}
\caption{Self-consistent scalar HF potential $S(x)$ at $T=0,p_f=0.2$ and $\gamma=0.2,0.6,1.0$ (from bottom to top), showing well resolved
baryons. Here and in Figs.~\ref{fig2}--\ref{fig4} only one spatial period of the periodic potentials is shown.}
\label{fig1}
\end{center}
\end{figure}
\begin{figure}
\begin{center}
\epsfig{file=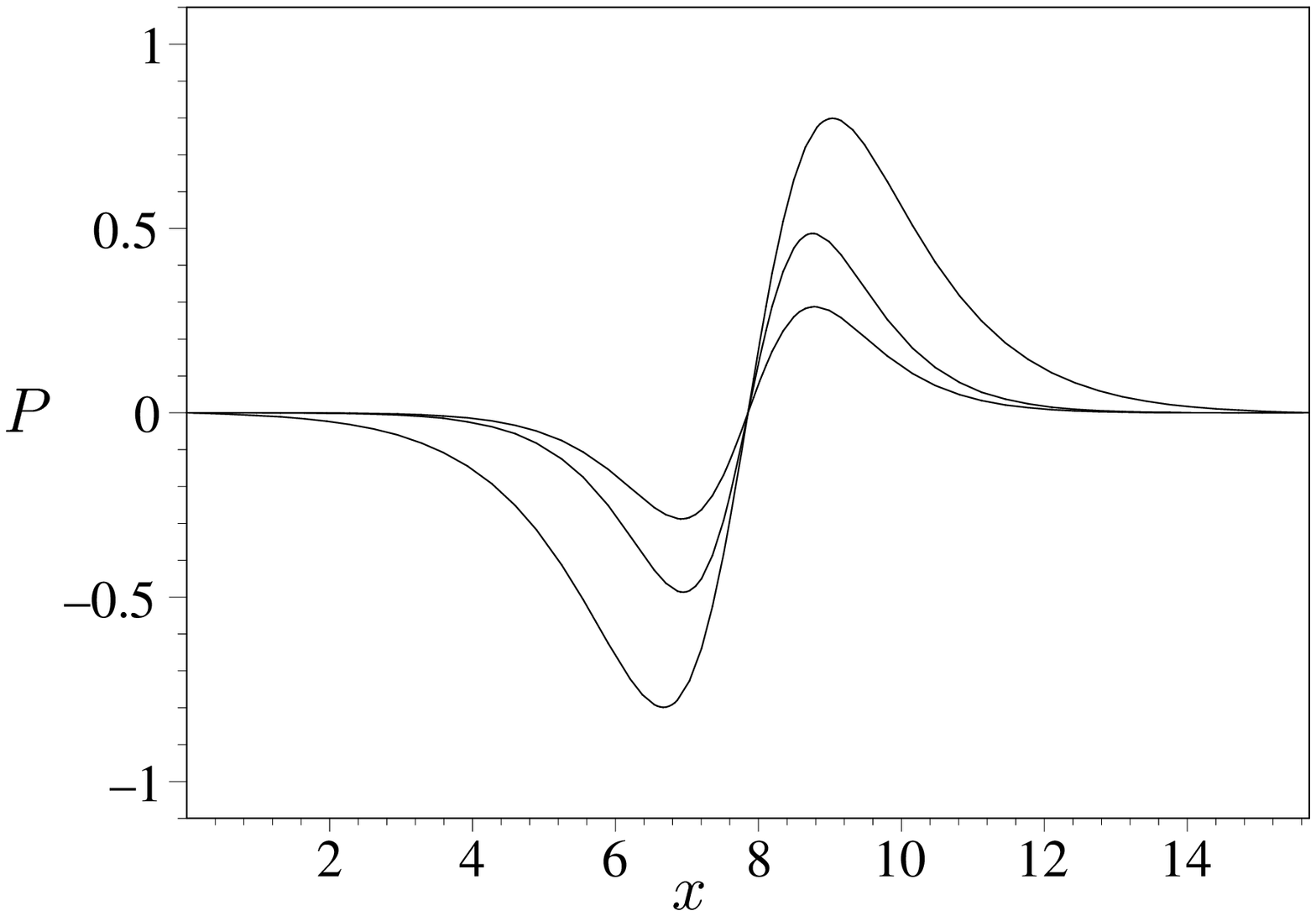,width=6.0cm,angle=270}
\caption{One period of self-consistent pseudoscalar HF potential $P(x)$ at $T=0,p_f=0.2$ and $\gamma=0.2,0.6,1.0$ (with decreasing amplitude).}
\label{fig2}
\end{center}
\end{figure}

\subsection{Numerical results}\label{sect2c}

We vary with respect to the Fourier components $S_{\ell},P_{\ell}$ which, owing to Eqs.~(\ref{A8}) and (\ref{A10}),
are real and satisfy $S_{-\ell}=S_{\ell},P_{-\ell}=-P_{\ell}$. The actual calculations were done as follows. 
We use $N=L/a=8$, i.e., perform single baryon computations with 8 different boundary conditions. 
In the sum over single particle energies we now have to subtract numbers of O(100 000), as compared to O(10 000) for a 
single baryon.
To keep computations feasible with MAPLE, we had to compromise on the size of the momentum space basis and choose
the smallest size which gave sufficient precision in the tests, $\bar{N}=50$ (corresponding to 201$\times$201 matrices). 
The total number of single particle states computed by diagonalization is therefore $8\times 201 = 1608$. We kept all
Fourier modes of $S,P$ up to $\ell_{\rm max}=6$, so that 14 real parameters had to be varied in total (the period $a$,
$S_0$, and $\left\{S_{\ell},P_{\ell}\right\}$ for $\ell=1...6$).
To test our MAPLE code, we computed the energy density of crystals where the analytic
solution is known (chiral spiral,  massless and massive GN models). In all of these cases the energy
density was reproduced correctly to 7 significant digits. The other uncertainty comes from the minimization procedure.
It was found that after only 20 conjugate gradient steps, the results as shown in the figures below did not change anymore
significantly. Under these conditions, all calculations could still be done using MAPLE on high-end PC's, without
need to switch to compiled programming languages. 

\begin{figure}
\begin{center}
\epsfig{file=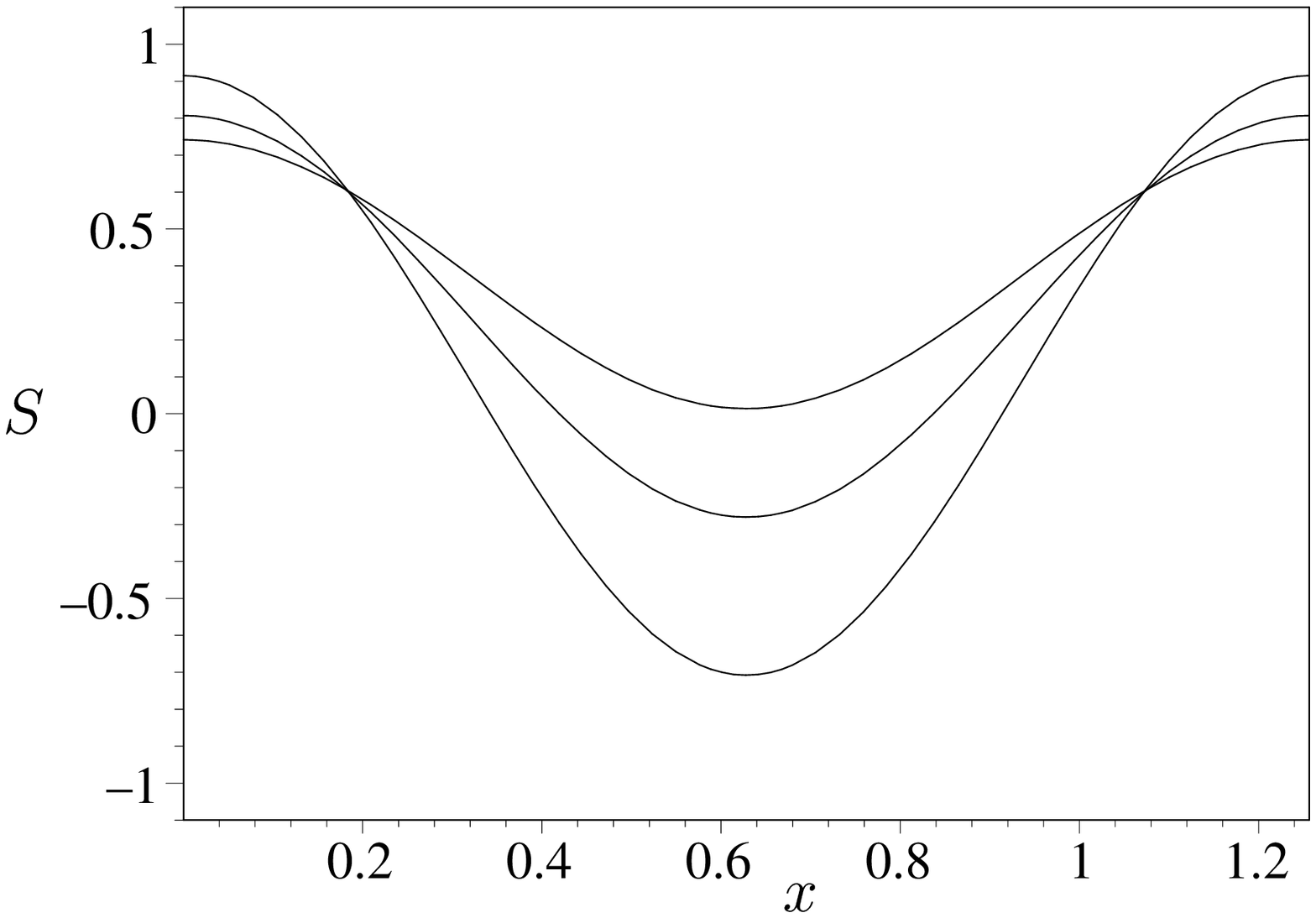,width=6.0cm,angle=270}
\caption{Same as Fig.~\ref{fig1}, but at $p_f=2.5$ where the baryons overlap strongly.}
\label{fig3}
\end{center}
\end{figure}

\begin{figure}
\begin{center}
\epsfig{file=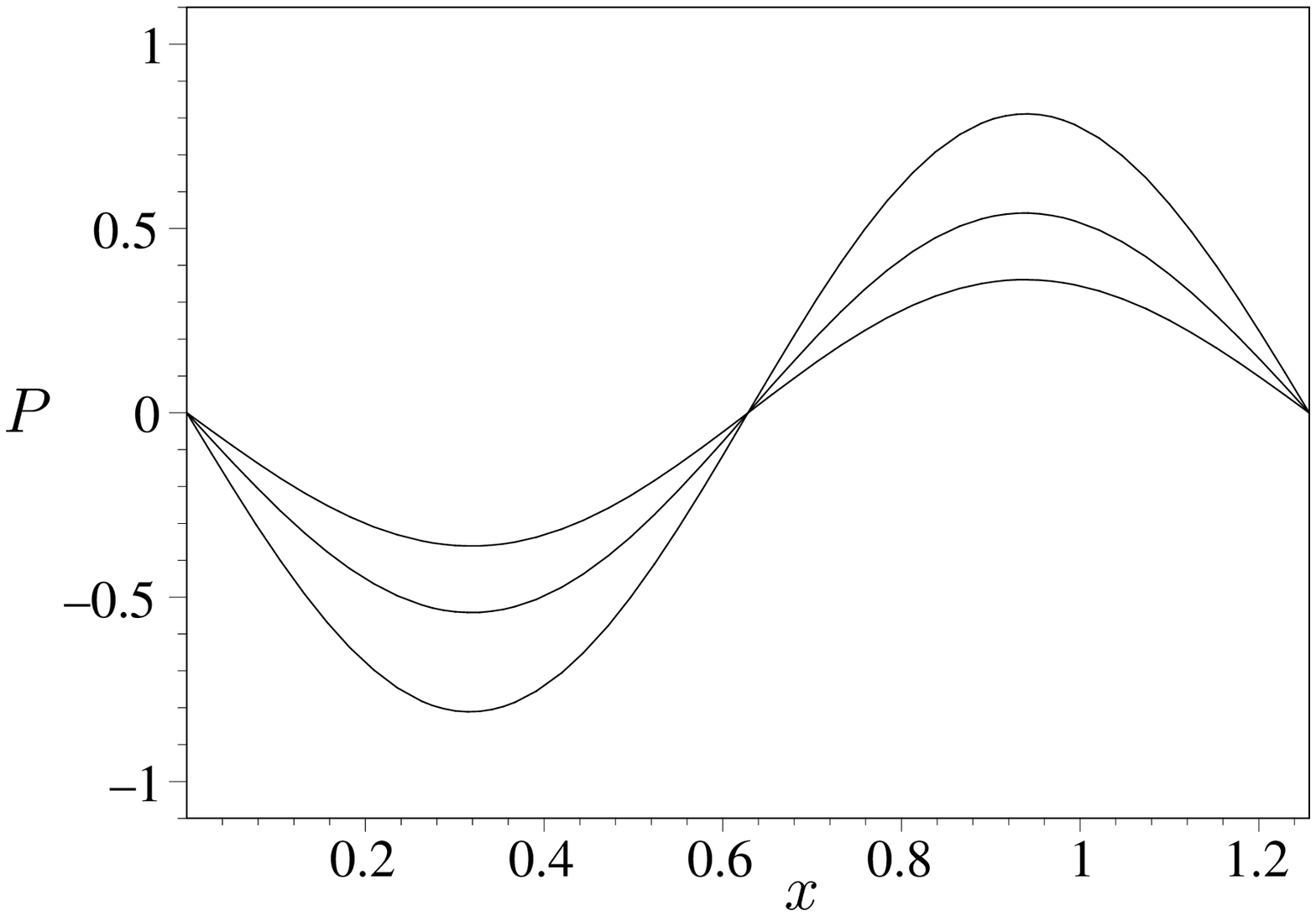,width=6.0cm,angle=270}
\caption{Same as Fig.~\ref{fig2}, but at $p_f=2.5$.}
\label{fig4}
\end{center}
\end{figure}

\begin{figure}
\begin{center}
\epsfig{file=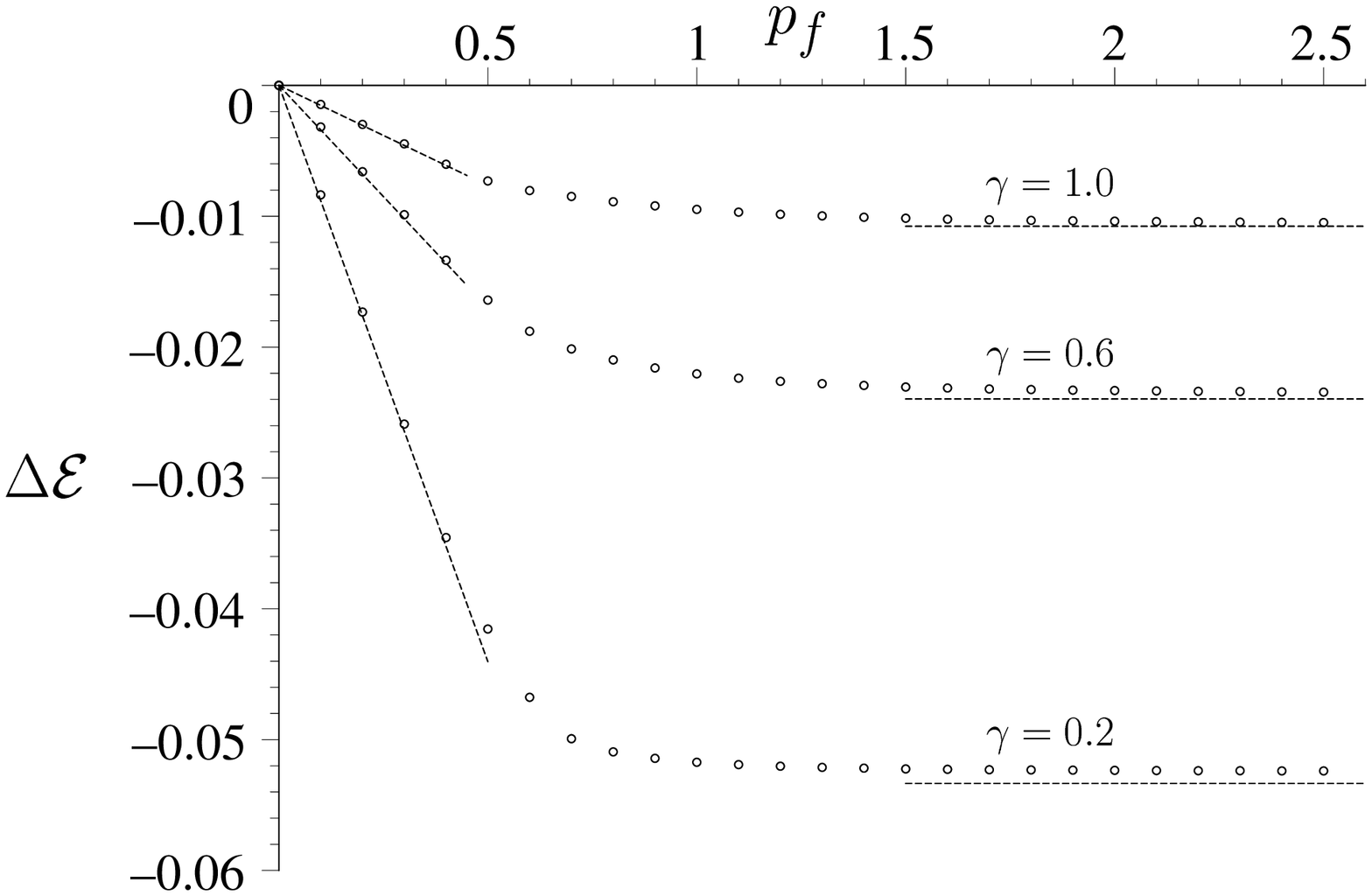,width=5.5cm,angle=270}
\caption{Difference between energy density of solitonic crystal phase and homogeneous phase versus $p_f$ for 3 different
values of $\gamma$. The straight line segments drawn show the analytical expectations for small and large $p_f$, respectively,
see the main text.}
\label{fig5}
\end{center}
\end{figure}

\begin{figure}
\begin{center}
\epsfig{file=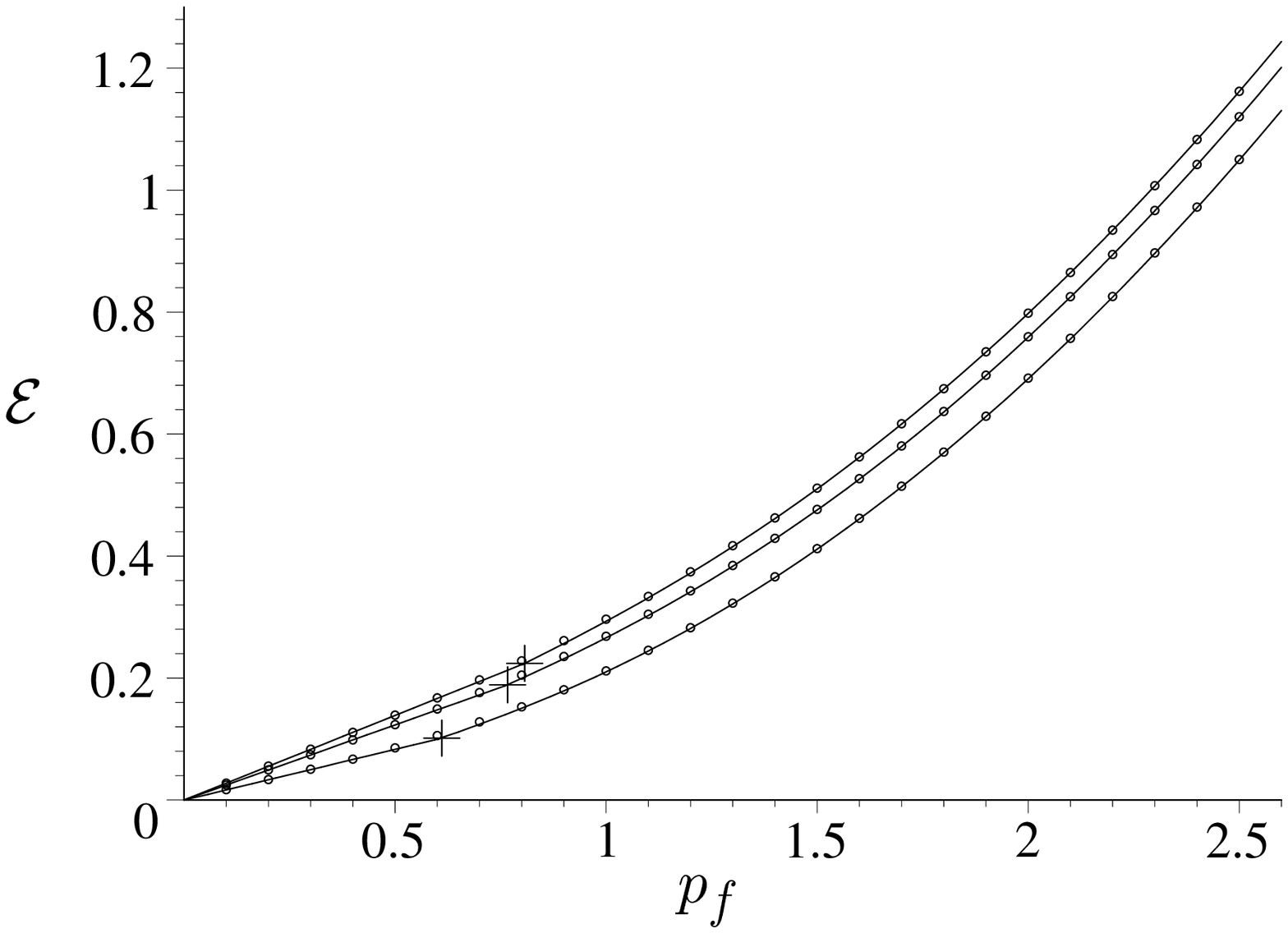,width=6.0cm,angle=270}
\caption{Ground state energy density of crystal at $T=0$ as a function of $p_f$ for $\gamma=0.2,0.6,1.0$ from bottom to top.
Points: numerical HF results, curves: asymptotic predictions according to Eqs.~(\ref{A12}) and (\ref{A22}), matched at the point marked
by a cross.}
\label{fig6}
\end{center}
\end{figure}

We now turn to the results of the $T=0$ computations.  Just as in the discrete chiral GN model, we found that it is always
energetically advantageous to let the Fermi surface coincide with the lower end of an energy gap, as expected from the 
Peierls effect. We first illustrate the self-consistent potentials which show no surprise. At low density, one recognizes the
shapes of clearly resolved individual baryons from Ref.~\cite{R15}, see Figs.~\ref{fig1} and \ref{fig2}.
At high density where the baryons overlap significantly, the lowest Fourier modes ($S_0,S_1,P_1$) dominate, as 
anticipated in our perturbative calculation (Figs.~\ref{fig3} and \ref{fig4}). Increasing $\gamma$ tends to wash out the oscillations at
all densities.
The energy difference between the crystal and the homogeneous phase 
is shown in Fig.~\ref{fig5} for 3 values of $\gamma$. 
As expected, the crystal phase is favored at all densities and $\gamma$ parameters. 
The horizontal lines at large $p_f$ show the asymptotic prediction of Eq.~(\ref{A21}), whereas the slopes of the 
straight lines near $p_f=0$ 
have been obtained from the baryon masses \cite{R15} and the mixed phase of the homogeneous calculation (see the appendix of \cite{R17}).  
This provides us with yet another useful test of the computations.
Fig.~\ref{fig6} shows the $p_f$-dependence of the energy density, now relative to the vacuum, for the same three values of $\gamma$.
The dots are numerical results. The curves have simply been obtained by matching the asymptotic expansions Eqs.~(\ref{A12}), (\ref{A22}),
at the point where they coincide (indicated by the cross).
At the scale of the figure, the agreement is perfect, reminiscent of similar findings in an earlier numerical study of the 
non-chiral GN model \cite{R18}.
\section{Constructing the phase diagram}\label{sect3}

\subsection{Grand canonical potential}\label{sect3a}

The phase diagram in the temperature-chemical potential plane is best analysed via the grand canonical 
potential density $\Psi$. The evaluation of $\Psi$ in the relativistic HF approach is well understood and follows earlier
studies of the non-chiral GN model, the only small complication being the fact that the spectrum is no longer
symmetric under $E\to -E$. The main building block is the familiar single particle contribution to $\Psi$, 
\begin{equation}
\Psi=- \frac{1}{\beta L} \sum_{\eta,n} \ln \left(1+{\rm e}^{-\beta(E_{\eta,n}-\mu)} \right).
\label{A23}
\end{equation}
For large positive or negative energy eigenvalues, one has to use perturbation theory in order to do the renormalization
analytically. The corresponding expression is 
\begin{equation}
\Psi_{\rm pert} = - \frac{2}{\beta}\int_{\bar{p}}^{\Lambda/2}\frac{{\rm d}p}{2\pi} \sum_{\eta} \ln 
\left(1+{\rm e}^{- \beta (E_{\eta,p}-\mu)}\right)
\label{A24}
\end{equation}
where $E_{\eta,p}$ denotes the 2nd order perturbative eigenvalue of the Dirac-HF Hamiltonian $H$. The standard manipulation
\begin{eqnarray}
\Psi_{\rm pert} & = & - \frac{2}{\beta} \int_{\bar{p}}^{\infty} \frac{{\rm d}p}{2\pi} \ln \left( 1+{\rm e}^{-\beta(E_{+1,p}-\mu)}\right)
\nonumber \\
& & - \frac{2}{\beta} \int_{\bar{p}}^{\infty} \frac{{\rm d}p}{2\pi} \ln \left( 1+{\rm e}^{\beta(E_{-1,p}-\mu)}\right)
\nonumber \\
& & + 2 \int_{\bar{p}}^{\Lambda/2} \frac{{\rm d}p}{2\pi} \left[ E_{-1,p}-\mu\right] 
\label{A25}
\end{eqnarray}
isolates the divergence in the sum over single particle energies, which can then be dealt with like at $T=0$, see Sec.~\ref{sect2}
and Ref.~\cite{R15}, adding the double counting correction and using the gap equation to eliminate unphysical parameters.
We then minimize $\Psi$ with respect to the potentials $S,P$. The result is the renormalized grand canonical potential density,
together with the self-consistent potential at a given temperature and chemical potential. A vacuum subtraction finally normalizes
$\Psi$ to 0 at the point ($T=0,\mu=0$) and removes remaining trivial divergences from the Dirac sea.

\subsection{Ginzburg-Landau theory}\label{sect3b}

There are regions in ($\gamma,\mu,T$)-space where a full HF calculation can be bypassed. 
This is the case whenever a microscopic GL theory can be derived, leading to an effective bosonic field
theory directly in terms of the scalar and pseudoscalar potentials $S,P$ with the fermions ``integrated out".
One can identify two such regions requiring somewhat different approximations. Close to the tricritical point at ($\gamma=0,
\mu=0,T={\rm e}^{\rm C}/\pi$), the potentials are both weak and slowly varying. This was exploited in Ref.~\cite{R16}, where a 
GL effective action was obtained analytically, using the derivative expansion around the free, massless
fermion theory.
The resulting effective action was then minimized by a numerical solution of the Euler-Lagrange equation, an inhomogeneous, 
complex non-linear Schr\"odinger equation. In this manner, a soliton crystal solution could be identified in a small region of
$(\gamma,\mu,T)$ space, separated by 2nd and 1st order transitions from a homogeneous massive Fermi gas phase. 
We refer the reader to this paper for more details. Another approximation allows
one to study the phase diagram for $\gamma\ll 1$ and $\mu \ll 1$, but without any restriction in temperature. 
Here, the potentials are still slowly varying but develop a large, constant scalar term $S_0$, i.e., a mass. 
The derivative expansion can still be trusted, provided one expands now around the {\em massive} free Dirac theory.
This technique was applied some time ago at $T=0$ to the baryons in the chiral
GN model near $\gamma=0$ \cite{R14}. The generalization to finite temperature and chemical potential is technically
rather involved. In particular, it does not lead anymore to analytic expressions as in Ref.~\cite{R14}, since the thermal integrals
with massive single particle energies cannot be done in closed form. As this technique was used here only for a small part of
the phase diagram, we refrain from giving all the details which have been worked out in Ref.~\cite{R19}. The resulting
effective action is a polynomial in $S,P$ and its derivatives with ($\gamma,\mu,T$)-dependent coefficients given in terms
of one-dimensional numerical integrals. It can be minimized numerically by varying the period and the Fourier coefficients of
$S$ and $P$, resulting in the equilibrium value of $\Psi$.
In this way, it is possible to extend the calculation of the 1st order transition line at small $\gamma$ down to zero temperature
and check that the base point of the critical line coincides with the baryon mass. Some examples of results for the phase boundary
thus obtained  will be shown below together with the results of the full HF calculation, see Sec.~\ref{sect3d} and Figs.~\ref{fig11},
\ref{fig12}.

\subsection{Perturbative 2nd order phase boundary}\label{sect3c}

As is well understood by now from similar studies of the non-chiral GN model or from the GL approach near the
tricritical point, the exact location of a contingent 2nd order phase boundary between crystal and homogeneous phases 
is a perturbative matter.
For this purpose, $S_0$ (i.e., the dynamical fermion mass) has to be treated exactly, whereas it is sufficient to keep $S_1,P_1$ from the
inhomogeneous terms and treat them in 2nd order almost degenerate perturbation theory (ADPT). As a matter of fact, right at the phase
boundary this amounts to naive 2nd order perturbation theory and a principal value prescription for integrating through the pole when
summing over single particle states \cite{R18}. The Hamiltonian is divided up according to
\begin{equation}
H=H_0+V
\label{A26}
\end{equation}
where 
\begin{eqnarray}
H_0 & = &  \gamma_5 \frac{1}{\rm i} \frac{\partial}{\partial x} + \gamma^0 m,
\nonumber \\
V & = & \gamma^0 2 S_1 \cos (2 p_f x) - {\rm i} \gamma^1 2 P_1 \sin(2 p_f x).
\label{A27}
\end{eqnarray}
To define the notation, we cast the unperturbed problem into the form ($\eta=\pm 1$ is the sign of the energy)
\begin{equation}
H_0 |\eta,p \rangle = \eta E |\eta, p \rangle, \qquad E=\sqrt{p^2+m^2}
\label{A28}
\end{equation}
with the free, massive spinors
\begin{equation}
\langle x |\eta,p \rangle = -\frac{{\rm i}\,{\rm sgn}(p)}{\sqrt{2L} E} \left( \begin{array}{c} {\rm i}p-m \\ \eta E \end{array} \right)
{\rm e}^{{\rm i}px}.
\label{A29}
\end{equation}
Matrix elements of $V$ are then given by
\begin{equation}
\langle \eta',p' |V|\eta, p \rangle  =  \frac{{\rm sgn}(p)\,{\rm sgn}(p')}{2 E E'} \left( {\cal A} S_1 + {\cal B} P_1 \right)
\label{A30}
\end{equation}
with  
\begin{eqnarray}
{\cal A} & = & \left[ \eta E({\rm i}p'+m)-\eta' E'({\rm i}p-m)\right]
\nonumber \\
& & \times (\delta_{p',p+2p_f}+\delta_{p',p-2p_f})
\nonumber \\
{\cal B} & = & {\rm i}\left[\eta \eta' E E' + ({\rm i}p-m)({\rm i}p'+m)\right]
\nonumber \\
& &  \times (\delta_{p',p+2p_f}-\delta_{p',p-2p_f}),
\label{A31}
\end{eqnarray}
leading to the following 2nd order energy shift,
\begin{equation}
\delta E_{\eta,p}  =  \frac{\eta(E^2 S_1^2 + p^2 P_1^2)+2 p_f E S_1 P_1}{(p^2-p_f^2)E}.
\label{A32}
\end{equation}
We insert 
\begin{equation}
E_{\eta,p} = \eta E + \delta E_{\eta,p}
\label{A33}
\end{equation}
into the single particle contribution to the grand canonical potential density,
\begin{eqnarray}
\Psi  &=&  - \frac{2}{\beta} \int_0^{\Lambda/2} \frac{{\rm d}p}{2\pi} \ln \left[ (1+{\rm e}^{-\beta(E_{1,p}-\mu)})
\right.
\nonumber \\
& & \left. (1+{\rm e}^{-\beta(E_{-1,p}-\mu)})\right],
\label{A34}
\end{eqnarray}
and linearize in $\delta E_{\eta,p}$.
Adding the usual HF double counting correction term and invoking the gap equation for the fermion mass at finite $T,\mu$
in the translationally invariant case, 
\begin{eqnarray}
0 & = & m(\gamma + \ln m)- \gamma 
\label{A35} \\
& & +  m \int_0^{\infty}\frac{{\rm d}p}{E}\left( \frac{1}{{\rm e}^{\beta(E-\mu)}+1}
+ \frac{1}{{\rm e}^{\beta(E+\mu)}+1}\right), 
\nonumber
\end{eqnarray}
to simplify the resulting expression, the perturbative correction to the grand canonical potential becomes
\begin{eqnarray}
\delta \Psi  & = & \frac{E_f^2 S_1^2+p_f^2 P_1^2}{\pi}\int_0^{\infty} \!\!\!\!\!\!\!\!\!\!\!- \ \  {\rm d}p
\frac{1}{E(p^2-p_f^2)}
\nonumber \\
& & 
\times \left(\frac{1}{{\rm e}^{\beta(E-\mu)}+1}+   \frac{1}{{\rm e}^{\beta(E+\mu)}+1}\right)
\nonumber \\
& & + \frac{2 p_f S_1 P_1}{\pi } \int_0^{\infty} \!\!\!\!\!\!\!\!\!\!\!- \ \  
{\rm d}p \frac{1}{p^2-p_f^2}
\label{A36} \\
& & \times  \left(\frac{1}{{\rm e}^{\beta(E-\mu)}+1}-   \frac{1}{{\rm e}^{\beta(E+\mu)}+1}\right)
\nonumber \\
& & + \frac{S_1^2+P_1^2}{\pi} \frac{\gamma}{m}- \frac{E_f^2S_1^2+p_f^2 P_1^2}{2\pi p_f E_f}\ln \left(\frac{E_f-p_f}
{E_f+p_f}\right).
\nonumber
\end{eqnarray}
The energies $E,E_f$ are defined with the mass $m=S_0$ and momenta $p,p_f$, respectively.
The principal value integrals are the only remnant of ADPT at the phase boundary \cite{R18} and
have to be evaluated numerically. The phase boundary can now be found using the following strategy. 
In 2nd order perturbation theory, according to Eq.~(\ref{A36}) we may write the grand canonical potential schematically as
\begin{equation}
\Psi = \Psi_{\rm hom} + {\cal M}_{11} S_1^2 + 2 {\cal M}_{12} S_1 P_1 + {\cal M}_{22} P_1^2
\label{A37}
\end{equation}
where all coefficients depend on  $m$ and $p_f$.
We have to vary $\Psi$ with respect to the 4 parameters, $m,S_1,P_1$ and $p_f$.
This yields the 4 equations
\begin{eqnarray}
0 & = & \frac{\partial \Psi_{\rm hom}}{\partial m} + S_1^2 \frac{\partial {\cal M}_{11}}{\partial m}
\nonumber \\
& & + 2 S_1 P_1 \frac{\partial {\cal M}_{12}}{\partial m} + P_1^2 \frac{\partial {\cal M}_{22}}{\partial m},
\label{A38} \\
0 & = & S_1 {\cal M}_{11} + P_1 {\cal M}_{12},
\label{A39} \\
0 & = & S_1 {\cal M}_{12} + P_1 {\cal M}_{22} ,
\label{A40} \\
0 & = & 
S_1^2 \frac{\partial {\cal M}_{11}}{\partial p_f} + 2 S_1 P_1 \frac{\partial {\cal M}_{12}}{\partial p_f} + 
P_1^2 \frac{\partial {\cal M}_{22}}{\partial p_f}.  
\label{A41}
\end{eqnarray}
At the phase boundary, Eq.~(\ref{A38}) can be simplified to the standard equation for the homogeneous
phase since $S_1,P_1$ vanish,
\begin{equation}
\frac{\partial \Psi_{\rm hom}}{\partial m} = 0.
\label{A42}
\end{equation}
Eqs.~(\ref{A39}), (\ref{A40}) represent a homogeneous system of equations which can be cast into the equivalent form
\begin{eqnarray}
{\rm det} {\cal M} & = & {\cal M}_{11}{\cal M}_{22}- {\cal M}_{12}^2 \ = \ 0,
\label{A43} \\
\frac{S_1}{P_1} & = & - \frac{{\cal M}_{12}}{{\cal M}_{11}}. 
\label{A44}
\end{eqnarray}
Dividing Eq.~(\ref{A41}) by $P_1^2$ and using Eqs.~(\ref{A43}), (\ref{A44}), we finally obtain the condition
\begin{equation}
\frac{\partial \,{\rm det}{\cal M}}{\partial p_f} = 0.
\label{A45}
\end{equation}
In order to determine the phase boundary, we have to find the points in the ($\mu,T$) plane where
Eqs.~(\ref{A42}), (\ref{A43}) and (\ref{A45}) hold simultaneously. Eq.~(\ref{A44}) then yields the unstable direction.
All of this can be done numerically to any desired accuracy. Before turning to the results,
it may be worthwhile to ask whether we can say anything about the outcome of the calculation beforehand. 
Indeed, it is easy to determine the asymptotic behaviour of the perturbative 2nd order sheet in the
limit $\mu\to \infty$, for any $\gamma$. Along the lines of a similar analysis in the appendix of Ref.~\cite{R18}
we arrive at the approximate expression for the grand canonical potential valid at large $\mu \approx p_f$,
\begin{eqnarray}
\Psi & = & \frac{S_0^2}{2\pi} [\gamma + \ln (2 p_f)]  - \frac{\gamma S_0}{\pi} 
+ \frac{2 X^2}{\pi} [\gamma +\ln (4 p_f)]
\label{A46} \\
&+ &  \frac{y^2}{4\pi} (2 \gamma - 1 +\ln y^2) - \frac{2}{\beta \pi} \int_0^{\infty}{\rm d}p \ln (1+{\rm e}^{-\beta \sqrt{p^2+y^2}})
\nonumber
\end{eqnarray} 
(using once again variables $X=(S_1+P_1)/2,y=S_1-P_1$).
$S_0$ and $X$ are not affected by finite temperature at all, so 
that Eqs.~(\ref{A17}) still hold. Minimization with respect to $y$ yields either $y=0$ (translationally invariant
solution) or the condition
\begin{equation}
\gamma + \ln y + 2 \int_0^{\infty} {\rm d}p \frac{1}{\sqrt{p^2+y^2}}\frac{1}{{\rm e}^{\beta \sqrt{p^2+y^2}}+1} = 0 .
\label{A47}
\end{equation}
In order to compute the phase boundary, we expand the integral for small $y$ \cite{R20},
\begin{equation}
\gamma + \ln y - \ln \frac{\beta y}{\pi} - {\rm C} + {\rm O}(y^2) = 0,
\label{A48}
\end{equation}  
where ${\rm C}$ is the Euler constant. The critical line where the non-trivial solution for $y$ disappears is 
then given by the following asymptotic expression valid at large $\mu$,
\begin{equation}
T_{\rm crit} = {\rm e}^{-\gamma} \left( \frac{{\rm e}^{\rm C}}{\pi}\right).
\label{A49}
\end{equation}
  
\begin{figure}
\begin{center}
\epsfig{file=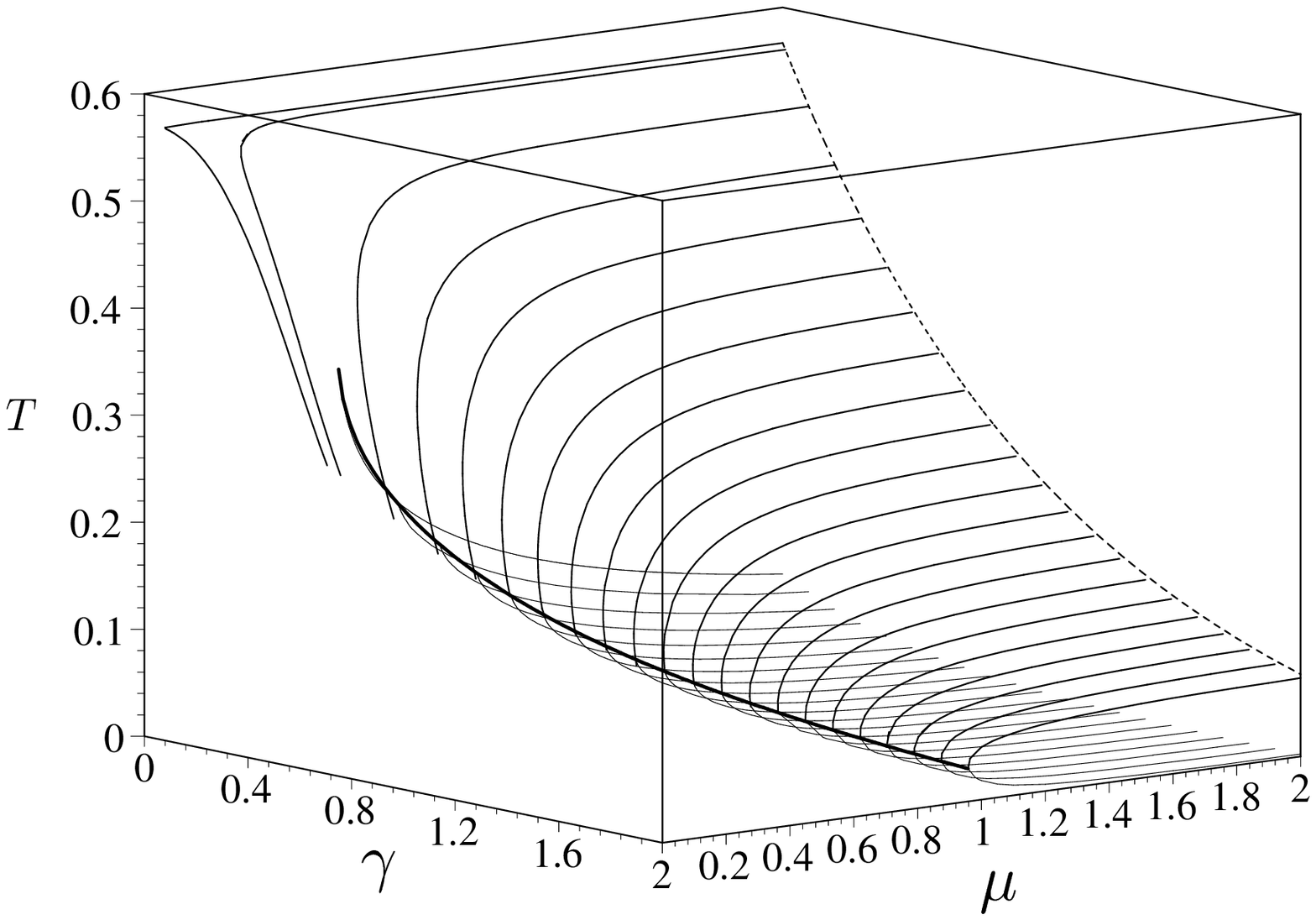,width=6.0cm,angle=270}
\caption{3d plot of the perturbative, 2nd order phase boundary in the chiral GN model (upper sheet), compared to the corresponding
phase boundary in the non-chiral GN model (lower sheet). The fat line is the tricritical curve of the latter model. The tricritical
curve of the chiral GN model cannot be determined by this  calculation. The dashed curve at $\mu=2$ is the asymptotic prediction 
of Eq.~(\ref{A49}).}
\label{fig7}
\end{center}
\end{figure}
Fig.~\ref{fig7} shows the results for the perturbative 2nd order sheet (a preliminary version of this plot has been given before in \cite{R21}).
This figure actually contains the 2nd order sheets for both the chiral and the non-chiral GN models to highlight 
the differences between the two models. The lower sheet ending at the fat black tricritical line belongs to the 
GN model with discrete chiral symmetry. To test our method, we have recalculated the curves shown here perturbatively. They agree
indeed with the results of Ref.~\cite{R22} where the same critical surface was deduced from the full, analytical solution of the HF problem.
The upper sheet in Fig.~\ref{fig7} is the new result for the chiral GN model. Here we have supplemented the equidistant curves at 
$\gamma=0.1,0.2, ... 2.0$ by 2 more curves at the small $\gamma$ values $0.01$ and $0.0001$.
This is useful to illustrate how this 2nd order sheet goes over into the horizontal critical line $T={\rm e}^{\rm C}/\pi$ in the chiral limit
$\gamma\to 0$.
We also compare the 2nd order sheet with the analytical prediction, Eq.~(\ref{A49}), at large $\mu$. As the dashed curve
shows, the full results are already indistinguishable from this formula at $\mu=2$. At low $\mu$, the curves 
bend over. Our calculation gives us no clue as to where the tricritical points are beyond which these curves turn
into 1st order critical lines. We will get back to this issue in the following
subsection when we discuss the full HF calculation. Notice also that at large $\gamma$, the perturbative sheets of both variants
of the GN model seem to come together at the same line (the tricritical line of the discrete chiral GN model), whereas this does not hold
anymore at small $\gamma$.

Finally, we should stress the fact that the sheet in Fig.~\ref{fig7} represents the surface where the homogeneous phase becomes
unstable towards crystallization in a continuous transition. If a 1st order transition occurs before reaching this sheet from the outside,
there will be no 2nd order transition and the corresponding part of the 2nd order sheet becomes obsolete. As
shown below, this is indeed what happens at sufficiently low temperatures.

\subsection{Non-perturbative 1st order phase boundary and full phase diagram}\label{sect3d}

\begin{figure}
\begin{center}
\epsfig{file=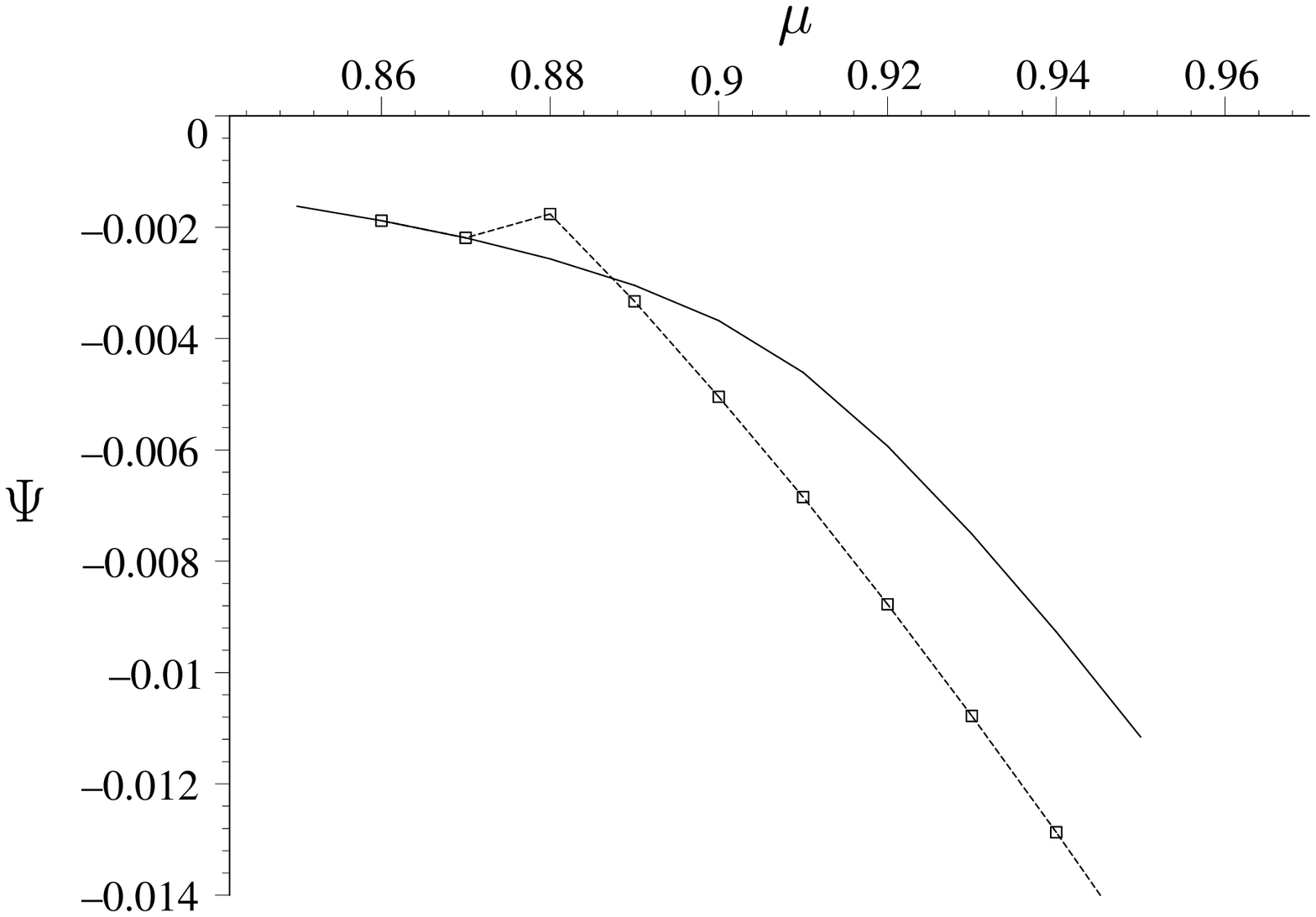,width=6.0cm,angle=270}
\caption{Determination of the 1st order phase boundary at $T=0.08,\gamma=1.0$. Points: Grand canonical potential from numerical
HF calculation vs. $\mu$. The crystal phase can be followed down to $\mu=0.88$. Solid line: Prediction assuming homogeneous
condensates only. The crossing of the 2 lines yields the critical chemical potential for a 1st order transition.}
\label{fig8}
\end{center}
\end{figure}

\begin{figure}
\begin{center}
\epsfig{file=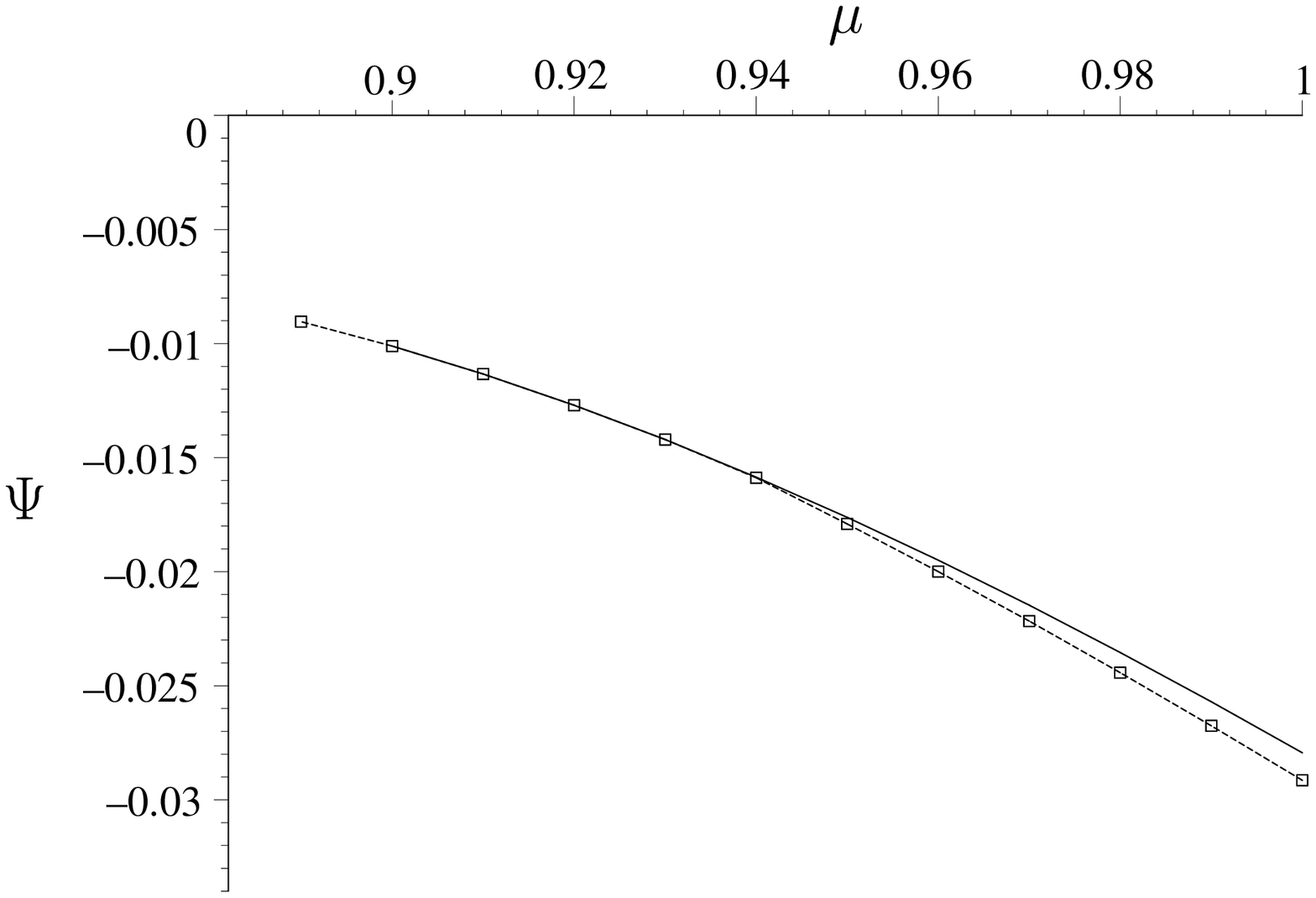,width=6.0cm,angle=270}
\caption{Same plot as in Fig.~\ref{fig8} at $T=0.12,\gamma=1.0$. The absence of line crossing is indicative of a continuous, 2nd order
phase transition.} 
\label{fig9}
\end{center}
\end{figure}

The most tedious task of the present study is to determine the 1st order phase boundary. 
For arbitrary $\gamma$ and $\mu$, no shortcut like GL theory is known and we have to resort to the full, numerical HF calculation. 
For a given point in the ($\gamma,\mu,T$) diagram,
we evaluate the renormalized grand canonical potential density $\Psi$ by minimization with respect to the period $a$ and the Fourier
components $S_{\ell},P_{\ell}$ of the mean field. The critical line is then constructed as follows. We evaluate 
$\Psi$ along a straight line trajectory for fixed $\gamma,T$ and several equidistant values of $\mu$, starting from inside the anticipated
crystal phase and proceeding towards lower $\mu$ values. We then plot $\Psi$ against $\mu$ and compare this thermodynamic potential
with the one of the homogeneous solution. In Fig.~\ref{fig8}, we illustrate the outcome of such a computation for the case
$\gamma=1.0,T=0.08$. The thermodynamically stable phase is the one with the lowest value of $\Psi$, hence the point of intersection 
of the 2 curves defines the critical chemical potential at this temperature.
Since we can follow the crystal solution beyond this point (before it jumps onto the other curve),
this is clearly a 1st order transition where two different solutions coexist at the phase boundary. The difference in slopes at the
intersection point translates into two different densities, so that a mixed phase would appear in a ($\rho,T$) phase diagram.
The critical point can be determined accurately in cases like that shown in Fig.~\ref{fig8}.
By contrast, Fig.~\ref{fig9} illustrates an example where the transition is likely to be 2nd order, namely at $\gamma=1.0,T=0.12$. 
Here, one does not see a crossing of the two curves. Due to the limited numerical accuracy, one cannot rule
out a very weak first order transition, therefore it is difficult to locate the tricritical point precisely in this manner.

\begin{figure}
\begin{center}
\epsfig{file=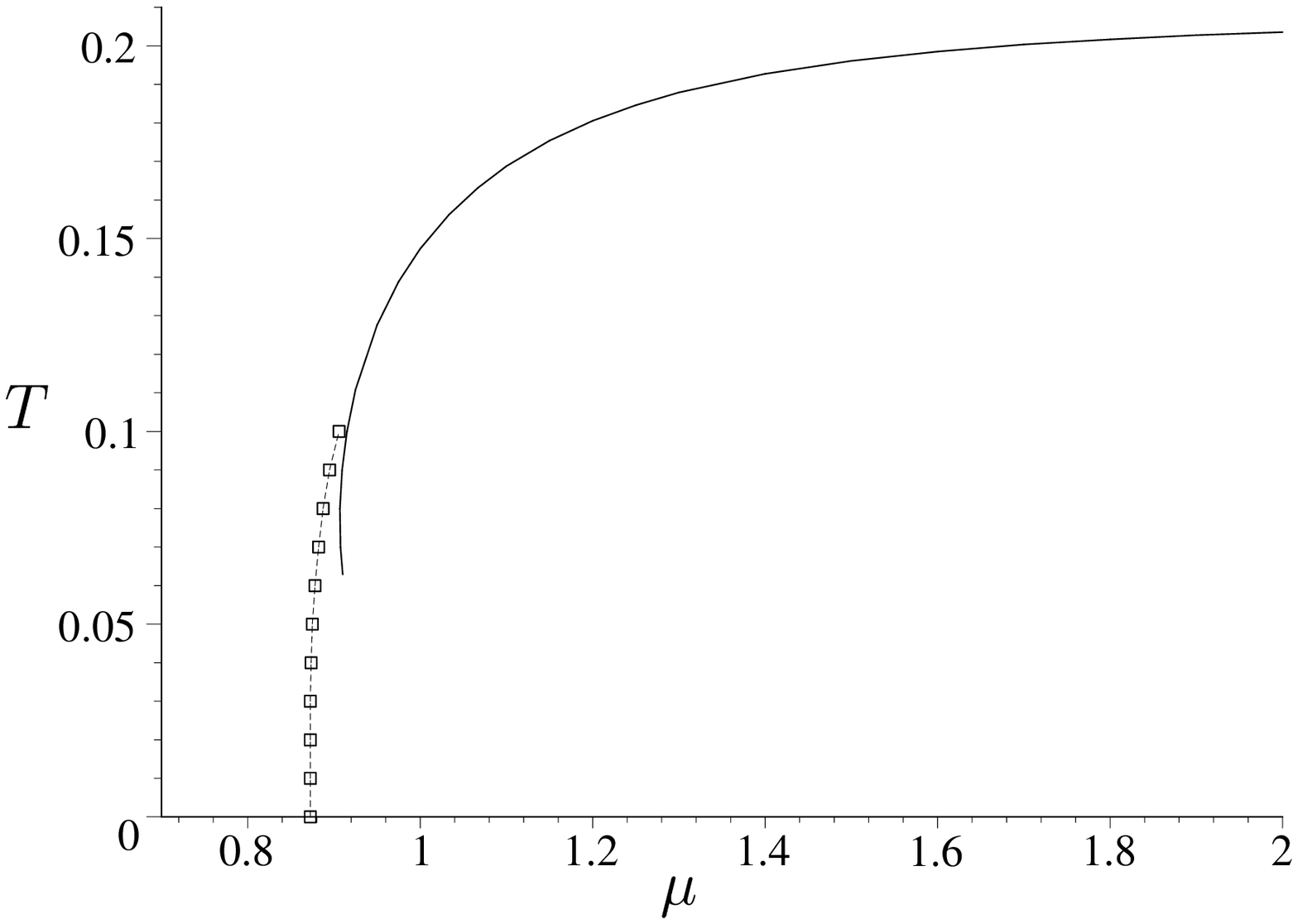,width=6.0cm,angle=270}
\caption{Example of a construction of the phase boundaries at $\gamma=1.0$. Solid line: 2nd order, perturbative critical line
from Fig.~\ref{fig7}. Points: 1st order, non-perturbative critical line determined as shown in Fig.~\ref{fig8}. The tricritical point has not yet
been located but must lie on the solid line, above $T=0.1$.}
\label{fig10}
\end{center}
\end{figure}

The result of such a computation of the 1st order critical line at $\gamma=1.0$ is shown in Fig.~\ref{fig10}. The solid line is the perturbative 2nd
order line from Sec.~(\ref{sect2c})  and Fig.~\ref{fig7}, without information on the tricritical point. The squares are numerically
determined 1st order phase transitions. We only show those points for which we could unambiguously identify a 
1st order transition.  Above $T=0.1$, there was no visible line crossing anymore. In this way, a small gap between the 2nd and
1st order phase boundaries is left. All we can say is that the tricritical point lies on the 2nd order line above the last 1st order
point shown, i.e., at $T>0.10$ in the case at hand. For as much as we can tell, the two critical lines are joined tangentially
at the tricritical point. Note that the base point of the 1st order line at $T=0$ drawn here is the baryon mass at $\gamma=1.0$ taken
from Ref.~\cite{R15}. The fact that the numerical points interpolate nicely between the baryon mass at $T=0$ and the perturbative phase 
boundary is a healthy sign, suggesting that the accuracy reached here is adequate. 
\begin{figure}
\begin{center}
\epsfig{file=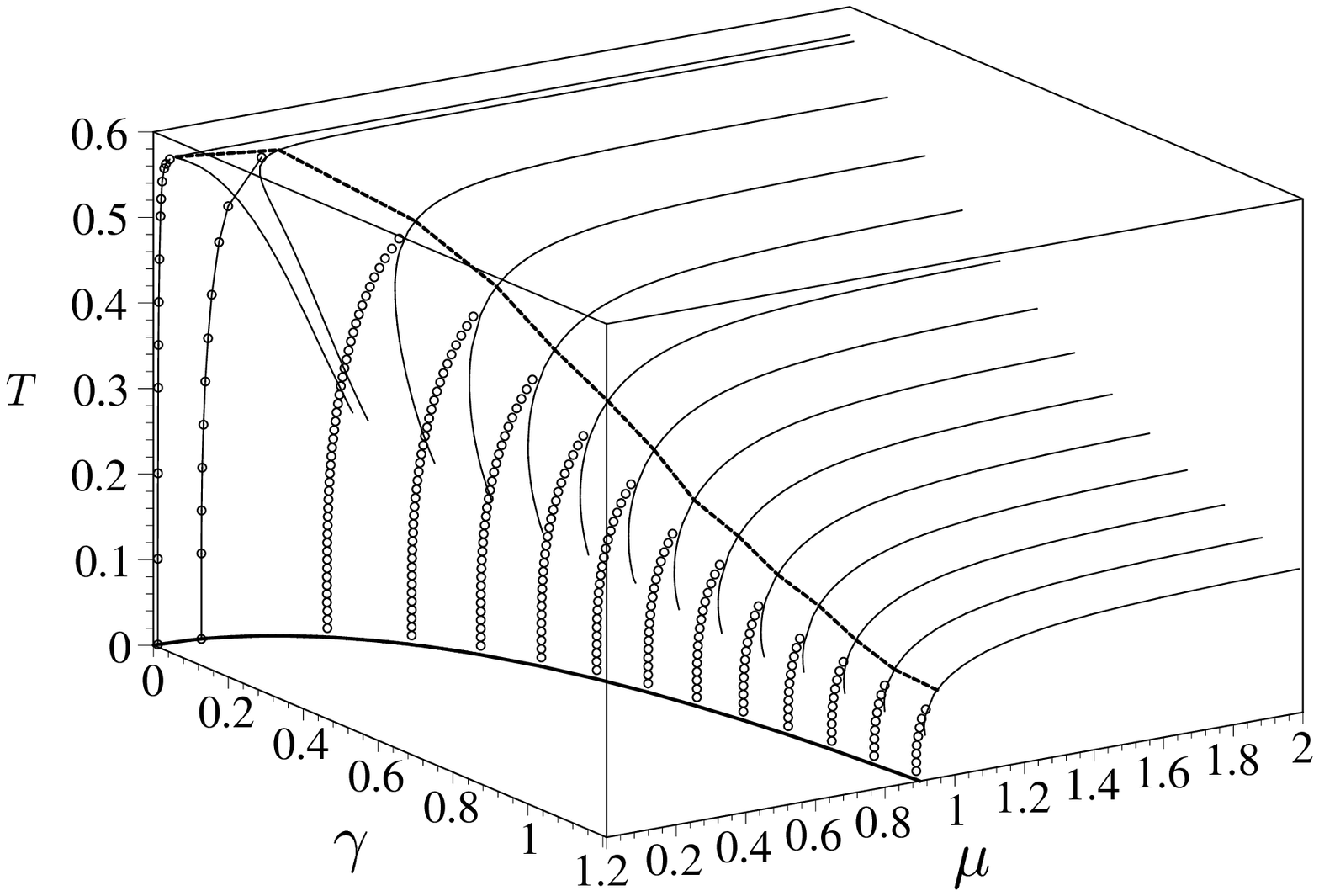,width=6.0cm,angle=270}
\caption{Summary of all the results about the phase diagram of the chiral GN model obtained in this work. Fat solid curve at $T=0$: Baryon 
mass, solid lines at fixed $\gamma$: Perturbative 2nd order sheet, points: numerically determined 1st order sheet, computed in steps of $\Delta \gamma=0.1,
\Delta T=0.01$. The 2 curves at very small $\gamma$ are taken from the GL analysis \cite{R19} and belong to  
$\gamma=0.01$ and $0.0001$, respectively. The fat line crossing the 2nd order sheet is the tricritical curve.}  
\label{fig11}
\end{center}
\end{figure}

In a lengthy numerical calculation with MAPLE, we have determined a number of 1st order critical lines, see Fig.~\ref{fig11}.
The solid curve at $T=0$ is the baryon mass from Ref.~\cite{R15}.  The thin lines are the 2nd order critical lines from Fig.~\ref{fig7}, the points are 
numerically determined 1st order transitions computed on a grid with resolution $\Delta \gamma=0.1, \Delta T=0.01$. 
Also shown are two additional curves at very small $\gamma$ (0.01 and 0.0001) obtained previously by means of the GL theory
\cite{R19}. These results confirm the picture discussed in connection with Fig.~\ref{fig10} and provide us with a first candidate for the
full phase diagram of the chiral GN model.
We find no indication of any further phase transitions beyond those which have been identified in the earlier study near the tricritical
point \cite{R16}.

\begin{figure}
\begin{center}
\epsfig{file=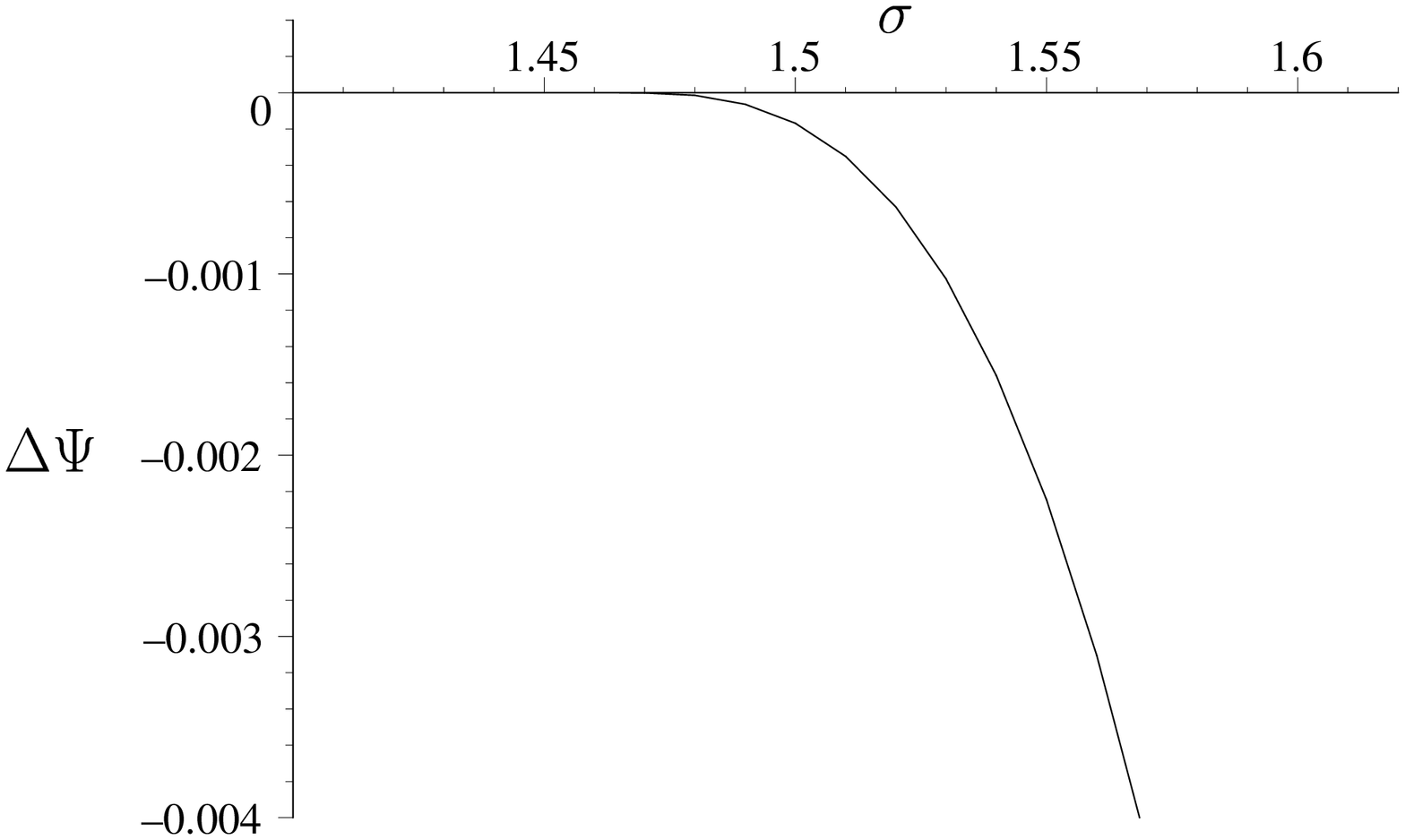,width=5.0cm,angle=270}
\caption{Difference in grand canonical potential between 2 phases near the tricritical point, using GL theory near $\gamma=0$. The rescaled
potential difference is plotted vs. $\sigma \sim (T_c-T)^{1/2}$ along the 2nd order critical line.}
\label{fig12}
\end{center}
\end{figure}
\begin{figure}
\begin{center}
\epsfig{file=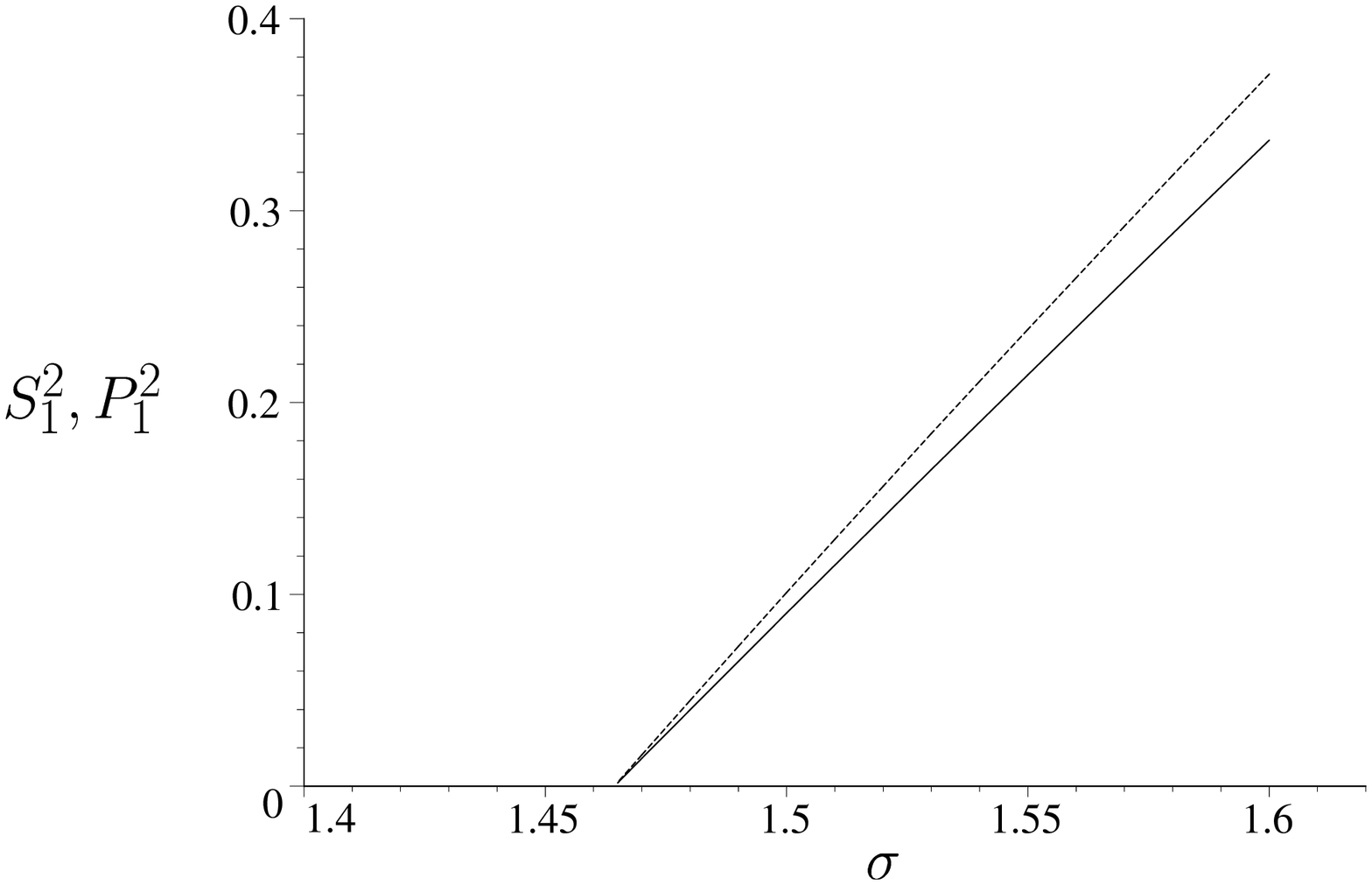,width=5.5cm,angle=270}
\caption{Square of rescaled Fourier amplitudes $S_1$ (solid) and $P_1$ (dashed) vs. $\sigma$ for the calculation
corresponding to Fig.~\ref{fig12}. The linear
behavior shows that $S_1,P_1 \sim (T_c-T)^{1/4}$ and locates the tricritical point precisely at $\sigma=1.464$.}
\label{fig13}
\end{center}
\end{figure}

In Fig.~\ref{fig11}, we have also plotted a tricritical line where the 1st and 2nd order critical sheets are joined together. As is clear from the gap
between the calculated 1st order sheet and this line, some extrapolation had to be used. We proceeded as follows. For a fixed
value of $\gamma$, we move along the 2nd order instability line, starting well below the expected tricritical point. We then perform the HF 
minimization and follow in particular the evolution of the largest Fourier components $S_1,P_1$. At the tricritical point, these are expected to vanish
with some power law $\sim (T_c-T)^{\alpha}$.
In order to find the relevant critical exponent $\alpha$, we went back to the
GL approach near $\gamma=0$ \cite{R16} and performed a similar analysis there. This has the advantage that we can work with a much
higher numerical precision in this regime. 
Let us first recall that to take advantage of simple scaling properties near the tricritical point, the variables $\mu,T$ have been
replaced by the rescaled variables 
\begin{equation}
\nu  =  2 \gamma^{-1/3}\mu ,\qquad \sigma= \sqrt{\frac{a}{T_c}} \gamma^{-1/3} \sqrt{T_c-T}
\label{A50}
\end{equation}
with $a=6.032,T_c=0.5669$ in Ref.~\cite{R16}. We now move along the perturbative phase boundary plotted in Fig.~6 of \cite{R16}
between $\sigma=1.4$ and $1.6$ enclosing the tricritical point. Along this trajectory the effective action is minimized with respect to the
Fourier components of $S_{\ell},P_{\ell}$ ($\ell \le 4$) and the period. The resulting grand canonical potential is compared to the
homogeneous calculation in Fig.~\ref{fig12}. Fig.~\ref{fig13} then shows clearly that the Fourier components $S_1,P_1$
vanish like $\sigma^{1/2}\sim (T_c-T)^{1/4}$. (Notice that the grand canonical potential and the Fourier components in Figs.~\ref{fig12}
and \ref{fig13} have been rescaled by the factors $2\pi a/\gamma$ and $\gamma^{-1/3}$, respectively, cf. Ref.~\cite{R16}.)
As a by-product, we have determined in this way a more accurate value of the tricritical point near $\gamma=0$ than in Ref.~\cite{R16}, 
namely $\sigma_t=1.464,\nu_t=3.039$. Coming back to the full HF calculation, we have located the point where $S_1,P_1$ vanish
along the 2nd order instability curve assuming the same critical exponent $\alpha=1/4$ for all $\gamma$.
Due to numerical limitations, the extrapolation is not as quantitative as in Fig.~\ref{fig13}, but still fairly straightforward.
The result is the tricritical curve drawn in Fig.~\ref{fig11}.

\section{Summary and conclusions}\label{sect4}

\begin{figure}
\begin{center}
\epsfig{file=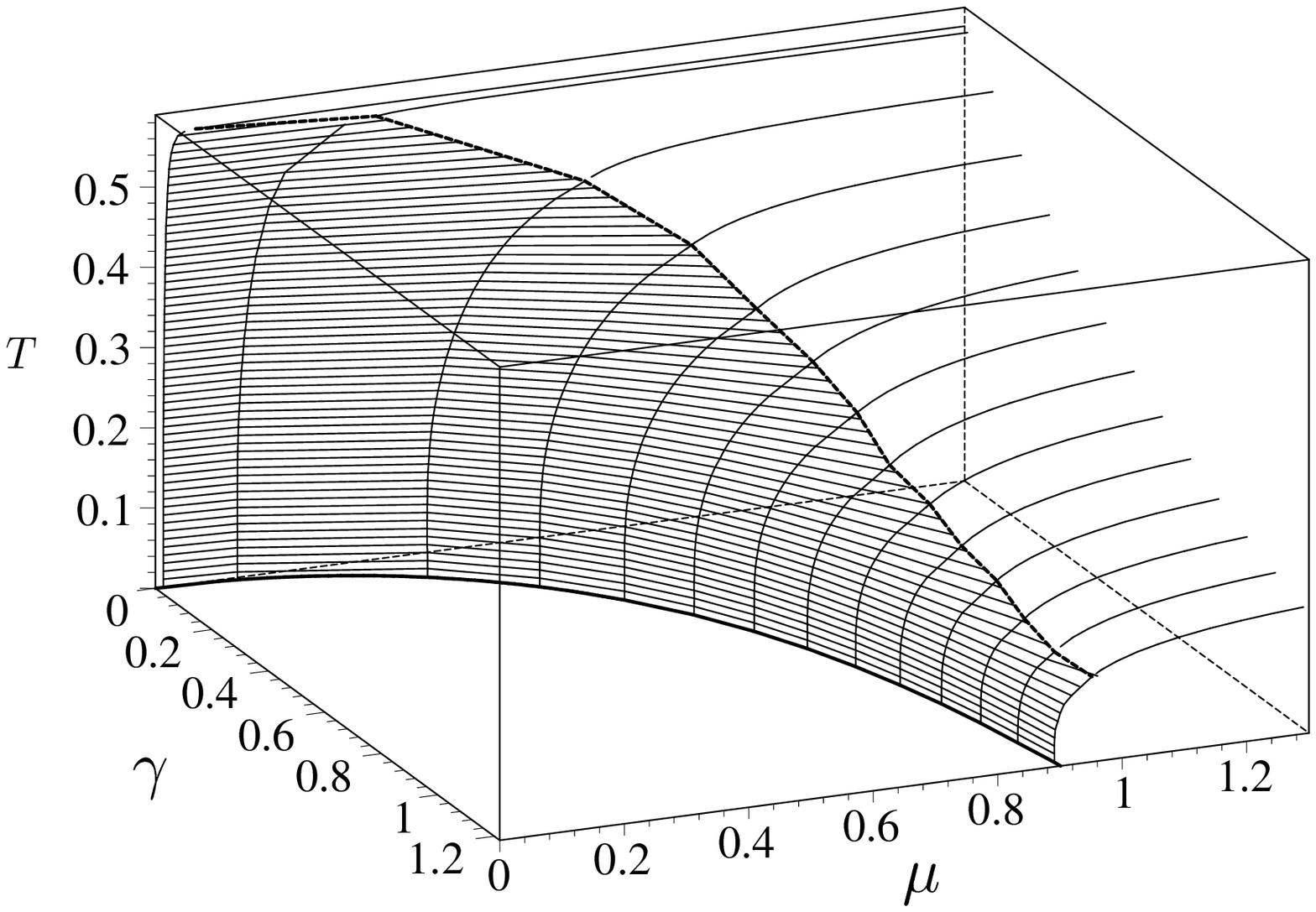,width=6.0cm,angle=270}
\caption{Phase diagram of the massive chiral GN model. The crystal phase with complex order parameter is separated from the
massive Fermi gas phase by 1st (dark shaded) and 2nd (light shaded) order critical sheets joined at a tricritical line.}
\label{fig14}
\end{center}
\end{figure}

To summarize, we have redrawn the phase diagram of Fig.~\ref{fig11} in a way which shows more clearly the shape of the 2 critical sheets,
see Fig.~\ref{fig14}. Here, we hide the ``engineering details" of the underlying construction still visible in Fig.~\ref{fig11}. Whereas the vertical
(1st order) 
lines have actually been computed via HF, the horizontal lines are composed of straight line segments joining neighboring points to guide
the eye.  The tricritical line separates 1st and 2nd order sheets. It is not completely smooth since this particular curve
is the most difficult part of the whole calculation, exceptionally sensitive to numerical inaccuracies. 

\begin{figure}
\begin{center}
\epsfig{file=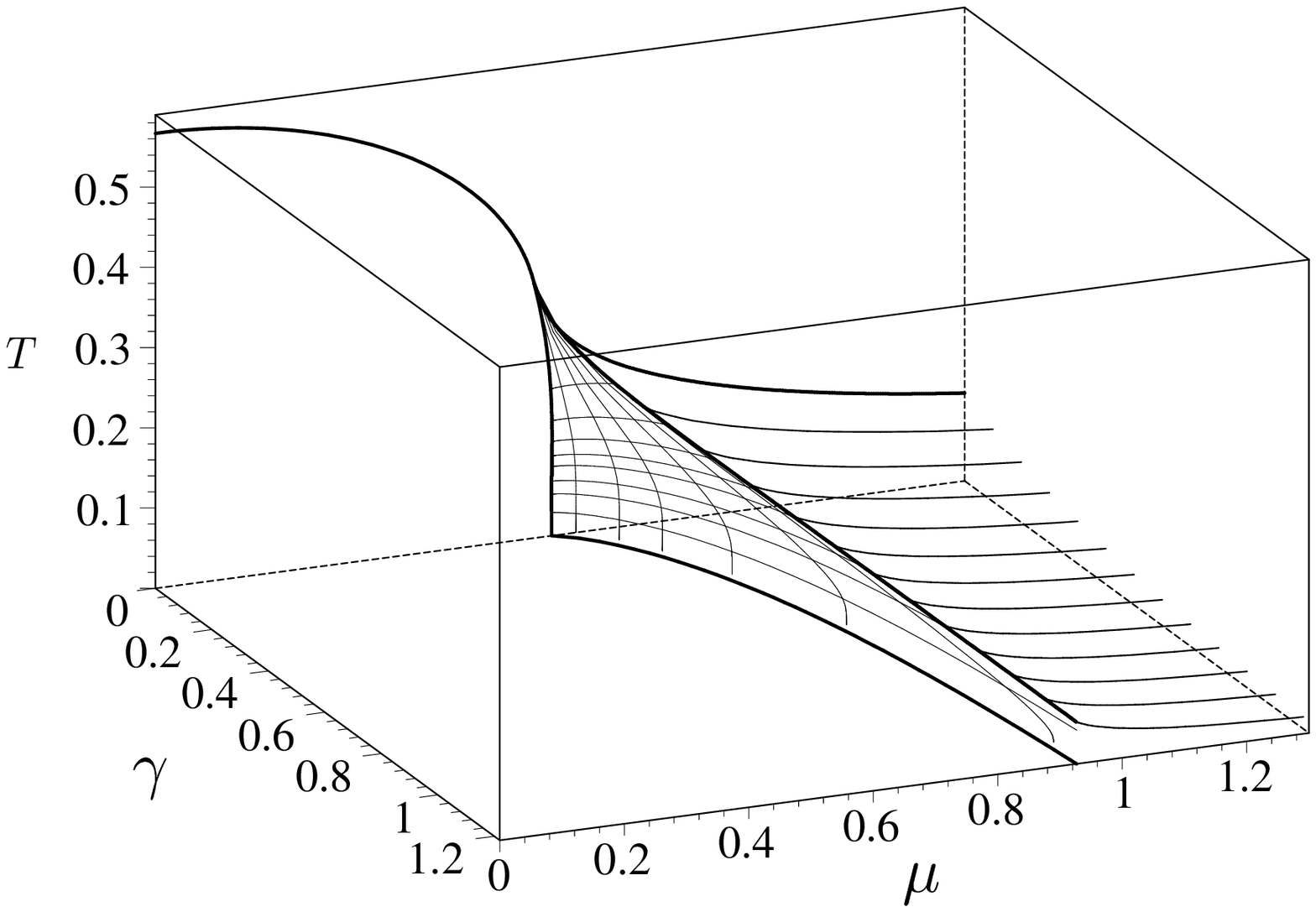,width=6.0cm,angle=270}
\caption{Phase diagram of the massive discrete chiral GN model, adapted from Ref.~\cite{R22}. The crystal phase with real order
parameter is separated by two 2nd order critical sheets from the massive Fermi gas phase, meeting at the tricritical line.}
\label{fig15}
\end{center}
\end{figure}

\begin{figure}
\begin{center}
\epsfig{file=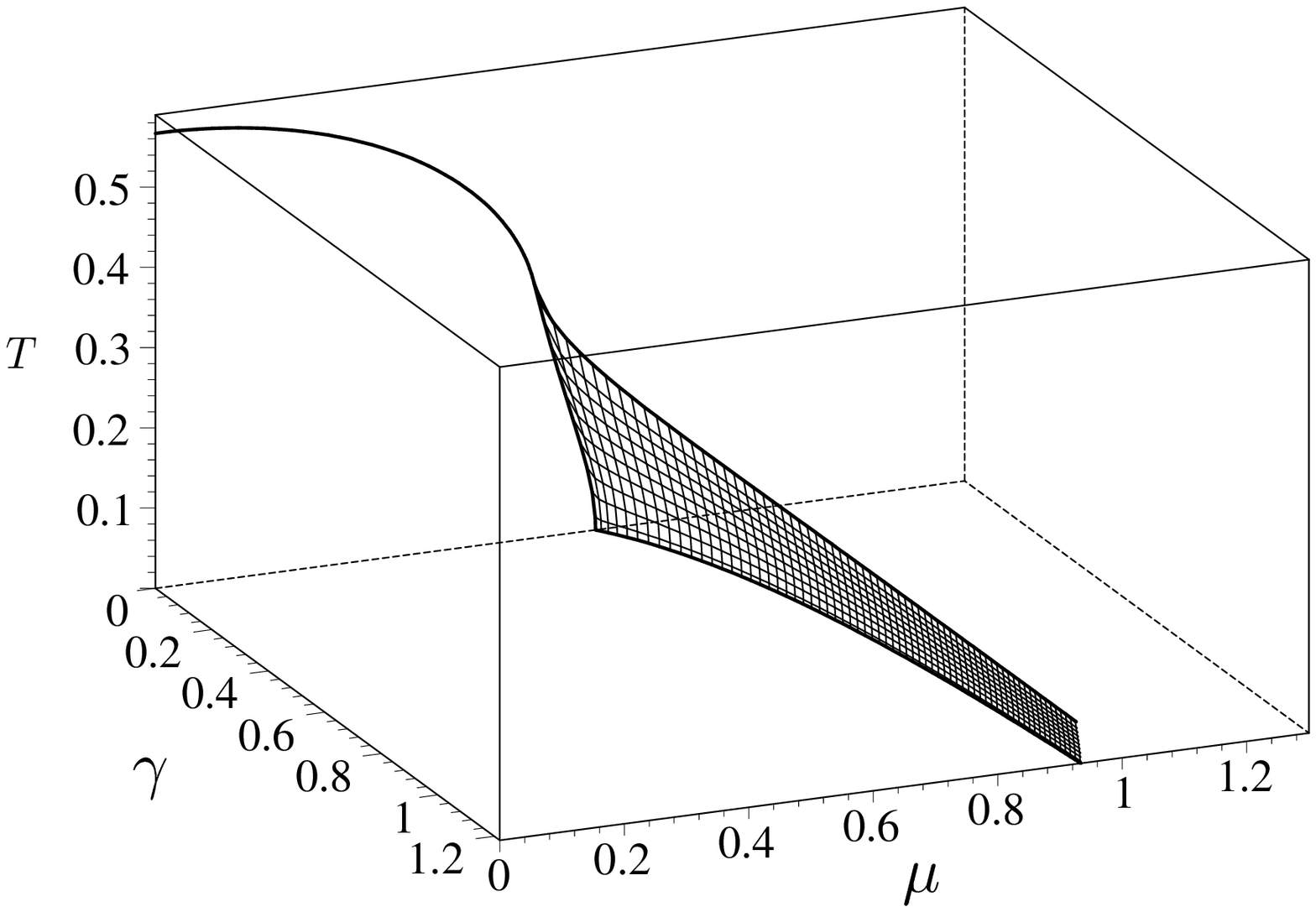,width=6.0cm,angle=270}
\caption{Common phase diagram for both variants of the massive GN model, assuming homogeneous condensates only. There is only 
a single massive Fermi gas phase, but the value of the mass changes discontinuously across the dark shaded 1st order sheet. 
The tricritical line agrees with the one in Fig.~\ref{fig15}. Adapted from Refs.~\cite{R12} and \cite{R22}.}
\label{fig16}
\end{center}
\end{figure}

It is interesting to compare this newly determined phase diagram of the massive chiral GN model to other related phase
diagrams. For this purpose, we have taken the results for the discrete chiral GN model from Ref.~\cite{R22} and plotted them at 
the same scale and under the same viewing angle as in Fig.~\ref{fig14}, cf. Fig.~\ref{fig15}. Here the 2 critical sheets are both 
2nd order and joined in a cusp rather than tangentially. The qualitative differences between Figs.~\ref{fig14} and \ref{fig15} are due
to the difference between continuous and discrete chiral symmetries of the two GN-type models, reflecting the corresponding 
universality classes. If one would only admit homogeneous phases as was done in the early works on these
phase diagrams, the 2 models would give identical results. This is illustrated in Fig.~\ref{fig16} adapted
from Ref.~\cite{R22}. Here, the dark shaded sheet
is 1st order, and there is only a single massive Fermi gas phase at $\gamma>0$.


Let us finally comment on some open questions. As pointed out above, to determine the tricritical line of the chiral
GN model requires some extrapolation of numerical results. The minimization becomes difficult close to the 
tricritical line where the effective potential is flat. Independent analytical work on the tricritical line and the 
critical exponent $\alpha$ discussed above would therefore be useful. One would also like to know the universality classes
to which the different variants of GN models belong. 

Another issue where further work is needed is related to the  
possibility of chiral twist in the massless case, recently discovered in Ref.~\cite{R11}. From the symmetry point of view, the situation
in the chiral limit may be characterized as follows: The Hamiltonian commutes with the generators ($P,Q,Q_5$) of translations
and vector/axial vector phase transformations of the fermions. A mass term (like in the vacuum or any homogeneous phase) breaks $Q_5$,
reducing chiral symmetry to U(1) vector transformations with the appearance of a massless Goldstone boson, the pion, and leaves $P$
unbroken.
The chiral spiral solution breaks $P$ and $Q_5$, but leaves the linear combination $P+\mu Q_5$ unbroken. Since one unbroken, continuous
symmetry is left, one expects only one gapless excitation, a mixture of a phonon and a pion. If the twisted kink crystal is realized, the
symmetry will be further broken down to one discrete combination of translation and $\gamma_5$ phase rotation, cf. Eq.~(\ref{A10a}). Such
crystals should feature two different gapless excitations, the phonon and the pion. 
In view of these different physics implications, it is important to reconsider the phase
diagram in the chiral limit once again and establish the thermodynamically most stable phases. 

In the massive chiral GN model, chiral symmetry is explicitly broken by the bare mass term, and $Q_5$ does not commute with $H$ anymore.
It is therefore unlikely that the $Q_5$ operator appears in a residual discrete symmetry of the condensates, as in the chirally twisted kink
crystal. 
The only remaining issue is then the fate of translational symmetry and its generator $P$. So far, we have tacitly assumed that translational
invariance breaks down to a discrete subgroup with concomitant periodic potentials, as is common in condensed matter systems. 
If this assumption would turn out to be wrong, the present calculation should be regarded as a variational calculation rather than the exact
solution of the model in the large $N$ limit, but such a scenario does not seem very likely to us.

\acknowledgements
One of the authors (M.T.) wishes to thank Gerald Dunne for helpful comments and discussions.

\end{document}